\documentclass{article}
\usepackage{newpxtext,newpxmath}
\usepackage{setspace}
\pdfoutput=1
\usepackage[utf8]{inputenc}
\usepackage[T1]{fontenc}
\usepackage[numbers,sort&compress]{natbib}
    \setcitestyle{authoryear,open={(},close={)}}
\usepackage{xcolor}
\usepackage{graphicx}
    \definecolor{defcolor}{rgb}{0,0,0.8}
\usepackage{xurl}
\usepackage[pdfauthor={A. Korcsak-Gorzo and C. Linssen et al.},pdftitle={Phenomenological modeling of diverse and heterogeneous synaptic dynamics at natural density},colorlinks=True,citecolor=defcolor,linkcolor=defcolor,urlcolor=blue]{hyperref}
\usepackage{enumitem}
\usepackage{csquotes}
\usepackage{amsmath}
\usepackage[capitalise]{cleveref}
\usepackage{caption}
\usepackage{geometry}
\crefname{enumi}{see Program}{see Programs}
\usepackage{fancyhdr}
\newcommand{\refnest}[1]{(\cref{#1})}

\title{Phenomenological modeling of diverse and heterogeneous synaptic dynamics at natural density}
\date{February 19, 2023}
\author{
Agnes Korcsak-Gorzo\textsuperscript{\textdagger, \textdaggerdbl, 1,2}, Charl Linssen\textsuperscript{\textdagger, \textdaggerdbl, 3,1}, Jasper Albers\textsuperscript{1,2},\\
Stefan Dasbach\textsuperscript{1,4}, Renato Duarte\textsuperscript{5,1}, Susanne Kunkel\textsuperscript{6,7}, \\
Abigail Morrison\textsuperscript{1,3,8}, Johanna Senk\textsuperscript{1}, Jonas Stapmanns\textsuperscript{1,2}, \\
Tom Tetzlaff\textsuperscript{1}, Markus Diesmann\textsuperscript{1,2,9}, Sacha J. van Albada\textsuperscript{1,10}\\
}
\begin{document}

\maketitle

\begin{footnotesize}
\textdagger~equal contribution

\textdaggerdbl~corresponding authors: a.korcsak-gorzo@fz-juelich.de (AKG), c.linssen@fz-juelich.de (CL)

\begin{enumerate}[itemsep=0mm]
    \item Institute of Neuroscience and Medicine (INM-6), Institute for Advanced Simulation (IAS-6), JARA-Institute Brain Structure-Function Relationships (INM-10), Jülich Research Centre, Jülich, Germany
    \item Department of Physics, Faculty 1, RWTH Aachen University, Aachen, Germany
    \item Simulation and Data Lab (SDL) Neuroscience, Institute for Advanced Simulation (IAS-6), \\Jülich Supercomputing Centre (JSC), Jülich Research Centre, Jülich, Germany
    \item Institute of Energy and Climate Research - Plasma Physics (IEK-4), \\ Jülich Research Centre, Jülich, Germany
    \item Donders Institute for Brain, Cognition and Behaviour, Radboud University, Nijmegen, Netherlands
    \item Neuromorphic Software Ecosystems (PGI-15), Jülich Research Centre, Jülich, Germany
    \item Faculty of Science and Technology, Norwegian University of Life Sciences, Ås, Norway
    \item Department of Computer Science 3 - Software Engineering, \\ RWTH Aachen University, Aachen, Germany
    \item Department of Psychiatry, Psychotherapy, and Psychosomatics, Medical School, \\ RWTH Aachen University, Aachen, Germany
    \item Institute of Zoology, University of Cologne, Cologne, Germany
\end{enumerate}

\end{footnotesize}

\newpage
\section*{Abstract}
    This chapter sheds light on the synaptic organization of the brain from the perspective of computational neuroscience.
    It provides an introductory overview on how to account for empirical data in mathematical models, implement such models in software, and perform simulations reflecting experiments.
    This path is demonstrated with respect to four key aspects of synaptic signaling: the connectivity of brain networks, synaptic transmission, synaptic plasticity, and the heterogeneity across synapses.
    Each step and aspect of the modeling and simulation workflow comes with its own challenges and pitfalls, which are highlighted and addressed.

\section*{Keywords}
    simulation, modeling, synaptic organization, spiking neural networks, plasticity, heterogeneity, phenomenological model, connectivity

\section{Introduction}

    Creating mathematical models from experimental neurophysiological data has grown into an established and essential method for investigating the brain.
    Based on these mathematical models and exploiting the upswing of affordable and powerful computing architectures over the last few decades, a new sub-field concerned with the computational modeling of neurobiological systems has emerged. The discipline using mathematical modeling and analysis methods to understand principles of brain organization, dynamics, and function is called \emph{computational neuroscience}. This discipline is also sometimes referred to as theoretical or mathematical neuroscience, each term having its own slightly different emphasis.
    One of the most challenging subjects of this comparatively young domain is the synaptic organization of the brain.
    This chapter reviews the status quo of synaptic modeling approaches. It is targeted primarily at experimentalists and aims to provide insight into ways data can be used to build mathematical or computational models.

    The methods described in the previous chapters of this volume reveal diverse dynamical processes and heterogeneous components involved in synaptic signaling at various spatial and temporal scales.
    This variability is amplified by the size and complexity of neurobiological systems: both the density and the total number of synapses in mammalian brains are impressive, the former being on the order of $10^9$ per cubic millimeter in the cerebral cortex \citep{Alonso-Nanclares2008} and the latter being estimated as roughly $5\times10^{14}$ in the human brain \citep{linden2018think}. Each cubic millimeter of the human brain contains on the order of $10^4-10^5$ neurons adding up to about $10^{11}$ neurons in the brain as a whole \citep{Azevedo2009equal}.

    \begin{figure}[!ht]
        \centering
        \includegraphics[width=.5\textwidth]{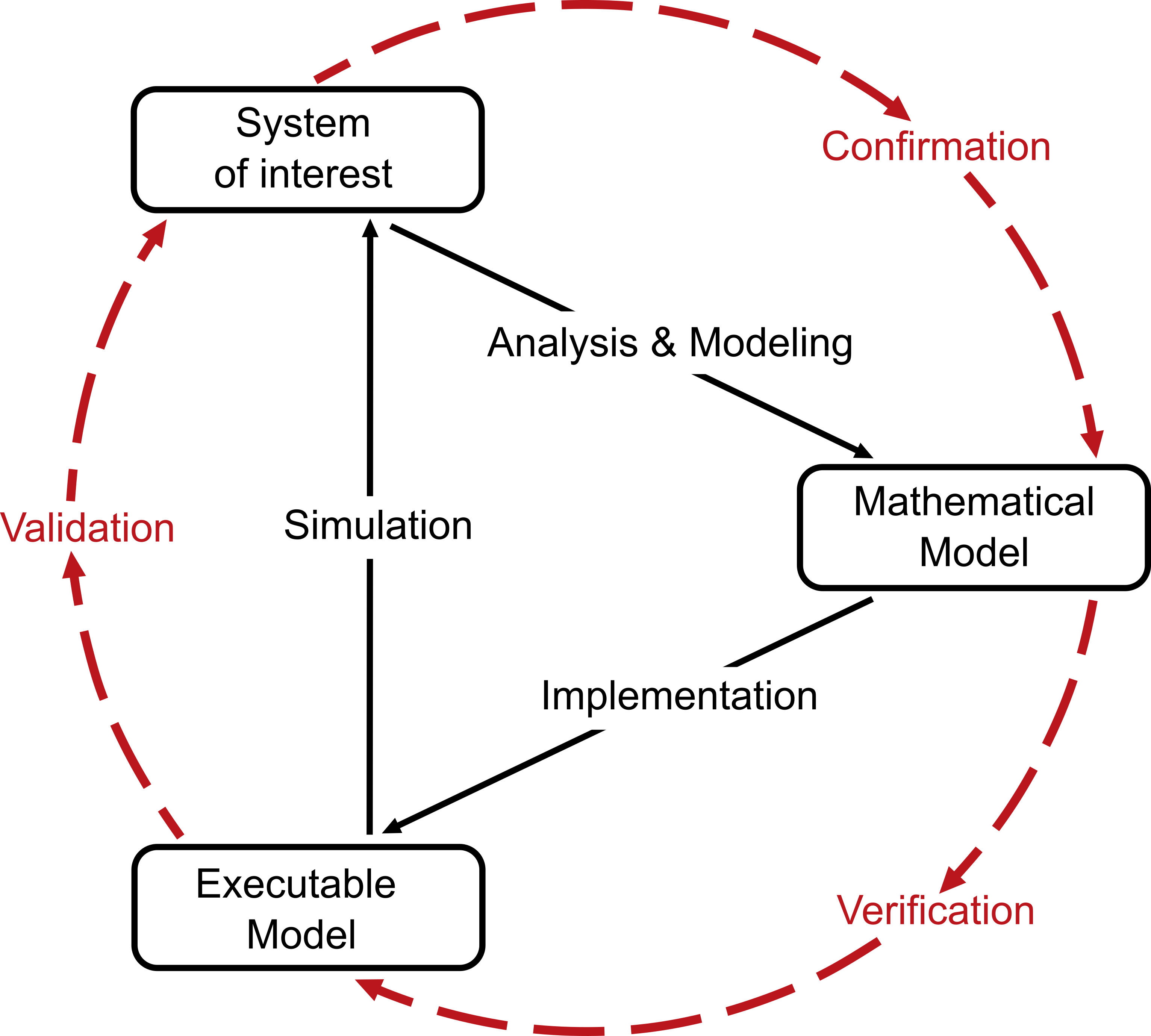}
        \caption{\textbf{Cycle of modeling and simulation}. The empirical data from the brain structure under study (the ``system of interest'') is first mathematically modeled and then implemented in software. Closing the loop, simulation results can be compared to experimental recordings. Reproduced with permission from Fig.~1 in \citet{10.3389/fninf.2018.00081}.}
        \label{fig:loop}
    \end{figure}

    How can we model such a heterogeneous, complex, and dense large-scale system?
    The process of modeling and simulation can be understood as a cycle, as depicted in \cref{fig:loop}.
    First, the experimental results recorded from the system of interest (here, the synaptic organization of the brain) are analyzed, and a mathematical model is formulated.
    Then, the mathematical model is translated into computer language, i.e., into an executable model that implements the mathematical operations needed to simulate the model.
    Finally, the model of the system is executed, whereby this simulation is the numerical analog to an experiment. This process yields results that can be compared with the experimental results. In turn, this comparison may deliver outcomes that can be used to improve the mathematical and computational models.
    Comparisons between the system of interest, the mathematical model, and the executable model ensure quality control. In general, three types of checks for correctness can be distinguished \citep{10.3389/fninf.2018.00081}: \emph{Confirmation} ensures that the mathematical formulation applies to the system of interest, \emph{verification} that the executable model sufficiently represents the mathematical model, and \emph{validation} that the simulation outcome is consistent with and predictive of the system of interest.
    This chapter focuses on the inner triangle of arrows: the practical methods to formulate mathematical models in an informed way, to translate them into manageable and correct executable models, to run simulations, and to inform further modeling choices using the obtained data.

    The challenge of computational neuroscience is to analyze the rich dynamics of neuronal systems and abstract their complexity into mathematical models that still capture essential characteristics of the experimental findings. At the same time, these models should be simple enough to be tractable and generalizable and thus reveal possible laws that govern the dynamical system.
    A fundamental question in this endeavor is what processes and variables are of interest and best describe the data.
    In synaptic organization, candidates include the connection strengths, synaptic time constants, delays, vesicle release characteristics, synaptic plasticity, and neuromodulation.

    Developing this thought further, a modeler needs to decide on the scale the model represents and how to parameterize it.
    It is advisable to constrain the number of model parameters to a minimal set that answers a specific research question. Limiting the parameter space increases the tractability, mechanistic interpretability, and robustness of the model and reduces the risk of overfitting.
    However, capturing biological detail and enhancing the direct link between parameters and their biological counterparts can usually only be done with a large set of parameters.
    Heterogeneity can be represented by introducing parameter value distributions, leading to additional parameters characterizing the dispersion and possibly higher-order properties of the corresponding distributions.
    Overall, a suitable parameterization involves a tradeoff between the model's controllability and biological plausibility.

    These decisions on which aspects of the system to express as variables and the choice of the corresponding model equations are abstraction steps: they formalize a hypothesis on which features are germane to the question at hand and which mathematical descriptions are appropriate for capturing the phenomena of interest (see \cref{sec:note_model_specifiction}).
    In general, this abstraction can be approached from two different directions.
    The \emph{bottom-up} approach starts from the low-level properties of the neurons and synapses making up the system and models the complexity step by step in the hope of achieving realistic dynamical and functional properties.
    However, one major point of modeling is to improve our understanding of a system. Given that the starting point is a poor understanding, this approach suffers from the fundamental problem that essential features might be abstracted away or obfuscated by an abundance of less relevant details. Another point can be to provide accurate predictions, even if we do not understand the model.
    The opposite approach, \emph{top-down} modeling, starts from the high-level dynamical, functional, or behavioral properties one would like to capture and then proposes concrete implementations. The drawback of this approach is that a model created in this way might not fully conform to biology, so it is difficult to draw conclusions about the brain.
    One solution is to use different degrees of abstraction at different scales to arrive at an understanding of the system, which is the motivation behind \emph{multi-scale modeling}.
    Ideally, biological realism is incrementally enhanced through cycles of data comparison and refinement (see \cref{fig:loop} and \cref{sec:note_model-refinements}).

    Some neurophysiological observations can be modeled with analytically solvable equations, i.e., in an exact way and usually with pen and paper.
    However, various simplifying assumptions generally flow into such abstractions, and deriving an analytical solution to a model's equations becomes less feasible as its complexity increases. For such cases, numerical solutions can provide a useful alternative.
    This computational approach tends to be slower, but it can validate the analytical approach by requiring fewer simplifying assumptions and it may even provide novel theoretical insight.

    Simple small network models frequently consist of equations that can be solved analytically or calculated numerically with few computational resources. However, both numerical and analytical approaches reach certain limits when attempting to replicate realistic neuron numbers in the volume of the brain region under consideration. As the number of neurons increases, the number of connections grows quadratically in networks without spatial dependence and linearly for distant neurons in models incorporating spatial dependence since most connections are local. To approximate \emph{natural density}, analytical techniques like mean-field theory sacrifice biological specificity. With sufficient computing power, numerical methods may solve model equations at natural density. A limitation is that, to date, this is only possible for small brains or small portions of larger brains.

    We restrict the scope of this chapter to phenomenological models, which represent the empirical relationship between phenomena without explaining the reason for the interaction. We neglect the molecular level or ultrastructure, i.e., structures visible at magnifications higher than that provided by standard optical light microscopy.
    Furthermore, we address spiking neuron models, mainly so-called point or few-compartment neurons, which neglect the precise morphology of the neuron, as the effective dynamics of a morphologically complex neuron can often already be meaningfully captured by such models.

    The models are usually translated to be executable by a computer, i.e., implemented in one of the various computer languages.
    One or an interconnected set of dynamical model components representing neurobiological entities such as neurons or synapses is \emph{simulated}, i.e., evolved in time, for a specific duration with a set of parameters, initial conditions, and stimuli.

    The generic framework that can numerically solve the dynamical equations of various models is called a \emph{simulator}.
    It solves complex interactions with many coupled differential equations typically incrementally in discrete time steps.
    Spiking neuronal network simulators also communicate spike events from senders to receivers.
    The event-based nature of synaptic interactions in the form of action potentials is advantageous for the efficiency of the simulation.
    Furthermore, a simulator can record dynamical state variables in the network or other observables like connectivity and apply stimuli during simulation.

    Since the simulator needs to organize and maintain the appropriate data structures in computer memory, a substantial amount of RAM may also be required, depending on the simulated neurobiological system.
    Biologically realistic network models are often large-scale to reflect or approach the natural density of neurons and synapses, have additional computational overhead due to heterogeneity, and are typically simulated for a long time, e.g., to gather statistics or study behavioral timescales.
    Thus, a simulator should be efficient and scale to high-performance architectures in terms of processing and memory usage.

    In addition to these performance aspects, criteria for a good simulator include functional completeness, numerical accuracy, and reproducibility of results. Furthermore, a simulator increases its value for the community if the available models are relevant to many members and the documentation is comprehensive and easy to understand.
    Developing a simulator that fulfills all these criteria and supports diverse models is complex and time-intensive. Consequently, simulators with peer-reviewed collections of implemented models are continuously developed as a community effort and shared as software packages to benefit the field of computational neuroscience.

    From the range of existing simulators for biologically inspired neuronal networks, this chapter focuses on NEST \citep{Gewaltig_07_11204}, an open-source software tool designed to simulate anywhere from small to large-scale networks of diverse spiking neuron models, and its associated domain-specific modeling language NESTML \citep{plotnikov2016nestml} that facilitates the creation of new neuron and synapse models. Other simulators with various scientific foci and special areas of application include NEURON \citep{hines2001neuron}, Brian \citep{stimberg2019brian}, Nengo \citep{bekolay2014nengo}, Arbor \citep{cumming2017arbor}, and ANNarchy \citep{vitay2015annarchy}. Some common (simulator-agnostic) interfaces are provided by PyNN \citep{davison2009pynn} and the modeling language NeuroML \citep{neuroml}.

    This chapter is structured as follows. Each section from \cref{sec:connectivity} to \cref{sec:heterogeneity} presents a two-step recipe to go from experimental data on a specific feature or mechanism of the synaptic organization to simulations: first, how to mathematically model experimental data, and second, how to simulate this model.
    \cref{sec:connectivity} starts with the most simplified view of brain circuitry, namely binary connections between neurons.
    This view is advanced in \cref{sec:synapses} to the notion of weighted connections.
    Then, dynamics is introduced to the existence of connections in \cref{sec:structural-plasticity} and the weights of those connections in \cref{sec:functional-plasticity}.
    Finally, \cref{sec:heterogeneity} discusses how to model the additional heterogeneity in all these features. Throughout the text, links to \cref{sec:notes} highlight aspects that are particularly challenging or contain pitfalls. Furthermore, links to \cref{sec:program-links} provide URLs to NEST code snippets that help the reader develop an intuition on how the usage of the discussed models would look in code. The chapter ends with some concluding remarks in \cref{sec:conclusions}.

\section{Connectivity} \label{sec:connectivity}

    \subsection{From empirical data to mathematical models}
        Anatomical and physiological experiments are yielding ever richer data sets on brain connectivity. Integration of these data into dynamical models can help gain insight into their implications for brain activity and function, besides identifying gaps in the data which can inform future experiments. The data cover diverse scales, ranging from electron microscopy at the sub-micron scale of synapses to light microscopy for neuronal morphology, paired recordings identifying fractions of connected neuron pairs, glutamate uncaging
        at the scale of tens to hundreds of microns, axonal tracing for long-range connectivity, and diffusion imaging for the whole-brain scale \citep{vanalbada2022bringing}.

        Despite the richness of the available data, none of these experimental approaches can specify full connectomes at the single-neuron resolution, especially in organisms with complex brain structures such as mammals. Therefore, we need to make predictions in order to complete the detailed connectivity data. One strategy is to find statistical regularities in the existing data and use these to extrapolate to missing data points. Of course, models do not need to be fully data-driven; various abstractions may be used to explore the influence of specific aspects of the connectivity. We illustrate the data-driven approach using the example of the cerebral cortex. In view of the incompleteness of the known cortical connectivity for any individual, we describe the connectivity in a probabilistic manner. A different strategy for generating the connectivity may be to grow connections according to developmental or other plasticity rules (for structural synaptic plasticity, see \cref{sec:structural-plasticity}).
    \begin{figure}[!t]
        \centering
        \includegraphics[width=\textwidth]{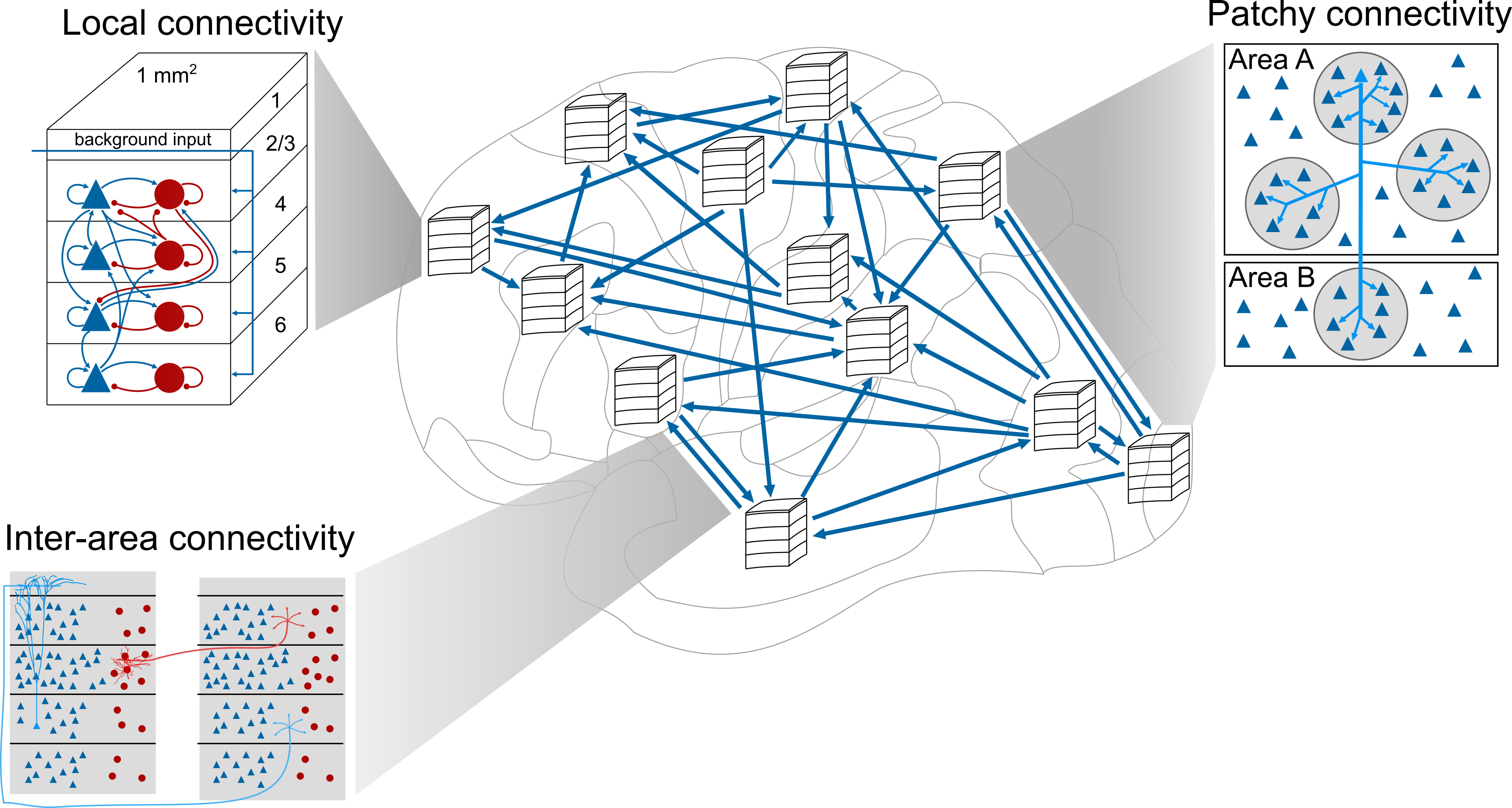}
        \caption{\textbf{Substructures of the cortex on different scales}.
        The neurons in the cortex are organized into areas on the macroscale (middle). In each area, neurons are organized into layers with distinct connectivity (here, shown as a microcircuit under $1~\textrm{mm}^{2}$ of cortical surface) and into populations of neurons with similar properties within each layer (upper left). Blue triangles and red circles represent excitatory and inhibitory neurons, respectively. Long-range projections connect the areas (lower left). Moreover, on an intermediate scale of millimeters, both intra-area and inter-area excitatory connections cluster into ``patches'' (right, only outgoing connections for one neuron shown). Adapted with permission from Fig.~1 in \citet{schmidt2018multi} under license \href{https://creativecommons.org/licenses/by/4.0/}{CC BY 4.0} originally from Fig.~1 in \citet{potjans2014cell} and Fig.~1 in \citet{kunkel2009simulating}.
        }
        \label{fig:scales}
    \end{figure}

        The cerebral cortex contains different types of excitatory and inhibitory neurons, distinguished by their morphology, electrophysiology, connectivity, and molecular make-up \citep{gouwens2019classification, hodge2019conserved}. We refer to the set of neurons of the same type in a given cortical area and layer as a \emph{population} (see \cref{fig:scales}). Connection probabilities are specific to both source and target populations. Both within and between areas, connectivity is also layer-specific. Furthermore, connection probability decays with the distance between neurons, both locally within areas and at longer ranges between areas \citep{packer2011dense, perin2011synaptic, ercsey2013predictive}. A further organizing principle is that excitatory connectivity tends to form patches \citep{Felleman91_1, Voges10_277}, meaning that neurons establish additional synapses onto other nearby neurons resulting in spatial clusters (see \cref{fig:scales}). 

        When formalizing these properties into models, a number of subtleties are involved \citep{Senk2022}. First, the term \emph{connection probability} needs to be defined carefully. This could, for instance, refer to either the total number of synapses divided by the product of the source and target population sizes or the probability for any neuron pair to be connected via at least one synapse. The two definitions diverge in the case of \emph{multapses}, multiple synapses between a given source and target neuron pair, often observed in reconstruction data \citep{kasthuri2015saturated}. Further, models can either allow self-connections, also called \emph{autapses}, or prohibit them. Moreover, beyond a certain model size, the spatial decay of the connection probability becomes important. To capture this, simulated neurons are assigned spatial coordinates, and additional specifications are necessary, including boundary conditions and the choice of connectivity profile. Common choices for the local profile are Gaussian and exponential functions, where the latter generally appears to be a better approximation to experimental data \citep{packer2011dense, perin2011synaptic}.

\begin{figure}[!t]
            \centering
            \includegraphics[width=\textwidth]{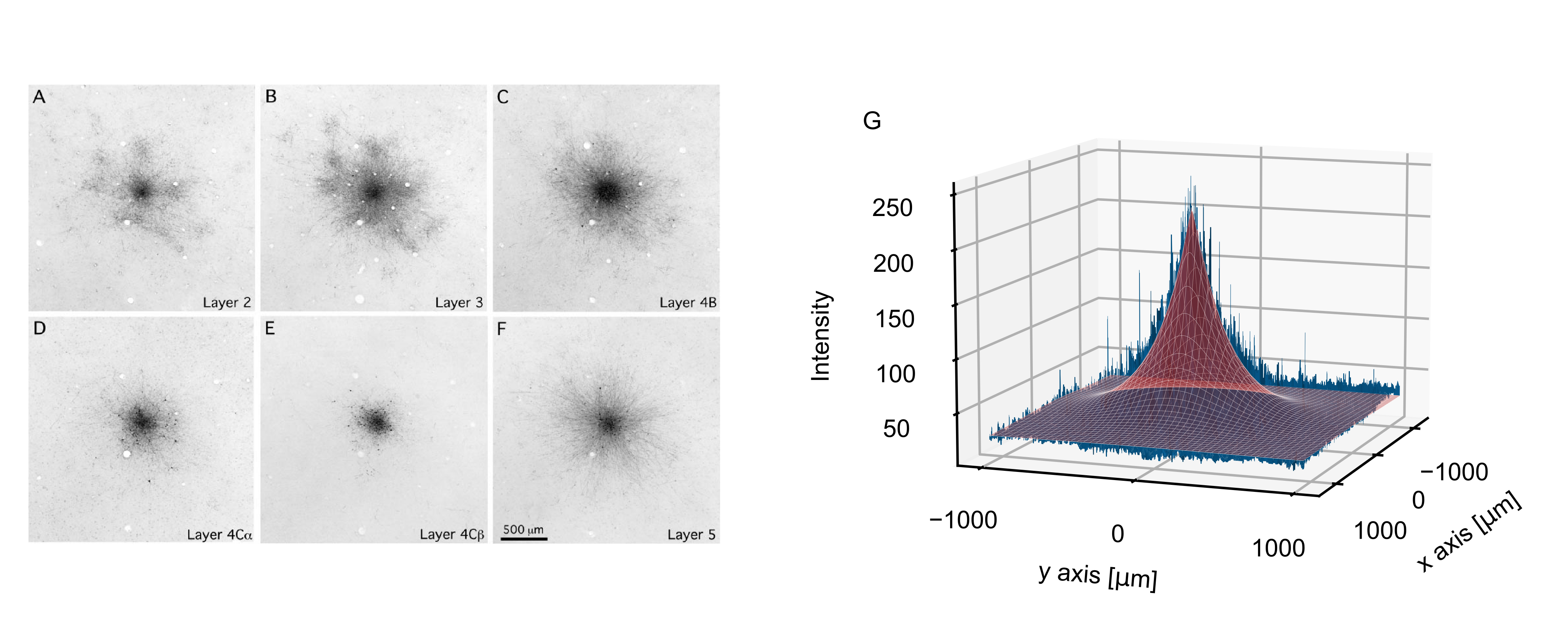}
            \caption{\textbf{Fitting a model of connectivity to observed data.} \textbf{A--F} Layer-resolved axonal tracing data from V1 of New World monkeys \citep[adapted from][Fig.~3, Copyright 2001 Society for Neuroscience]{Sincich4416}. Images show staining of cortical layers after an injection of biocytin into layer 3, anterogradely staining axons. \textbf{G} Fit of a 2D symmetric exponential function $f(r)=a e^{-\lambda/r}+b$ (red) to the axonal density distribution (blue) of layer 5. The model does not fully capture the connectivity profile for layers 2, 3, and 4B (A, B, C), which display patchy connectivity.}
            \label{fig:connectivity}
        \end{figure}

        \cref{fig:connectivity}A--F illustrates the local decay of connectivity with distance. Choosing a symmetric exponential as a model, \cref{fig:connectivity}G shows that fitting to the experimental data can reveal fundamental constants such as the characteristic length $\lambda$.

        When including \emph{patchy connectivity}, the spatial position of the patches can be specified via a radial distance from a cell body and an angle \citep{Voges10_277}. Further possible parameters are the number of patches, the size of each patch, and the degree of overlap between patches. Layer-specific axonal tracing data, such as fractions of supragranular labeled neurons from retrograde tracing experiments \citep{markov2014anatomy}, can inform the laminar inter-area patterns of cortical models. Here one should pay attention to the fact that, on the target side, axonal tracing tells us about axonal or synaptic locations but not about the locations of the target cell bodies. To a reasonable approximation, one can statistically estimate which synapses are established on which target neurons using morphological reconstructions, a method that assigns the number of synapses proportionally to the total length of dendritic elements in the vicinity of the synapses \citep{rees2017weighing, schmidt2018multi}.

        This is only a tiny selection of data and features that can be included in neuronal network models. One can go into greater complexity and, for example, consider the higher-level organization of networks, such as hierarchical modularity or small-world properties. For a further discussion on model detail in general, see \cref{sec:conclusions}.

    \subsection{From mathematical models to simulation}\label{subsec:conn2}

        To simulate how the dynamics of a neuronal network model evolve, the mathematical model description needs to be translated into an executable one. This is preferably done using a dedicated simulator to avoid mistakes in the implementation and to enhance comparability and reproducibility of results (for the precise definition of the different forms of \emph{reproducibility}, see \cref{sec:note_reproducibility}.)
        Executing a neuronal network simulation typically involves two successive phases:
        during the \textit{build phase}, the network is set up on the machine by instantiating objects and data structures for neurons and synapses.
        The subsequent \textit{simulation phase} propagates the network state for a specified biological model time.
        How fast a simulation runs, i.e., how the biological model time relates to the wall clock time, crucially depends not only on the machine specifications but also on the representation of the network model on the machine. Parallel computing combines the computational power of many separate compute cores or nodes to enable large-scale simulations; to this end, NEST uses a hybrid approach with the Message Passing Interface (MPI) and Open Multi-Processing (OpenMP). The former enables parallel computing on multiple processors with distributed memory, while the latter enables parallel computing even on single processors with shared memory, referred to as \emph{threading}. The total number of so-called \emph{virtual processes} is determined as the product of the number of MPI processes and the number of OpenMP threads per process.
        A direct mapping between network structure and hardware is in general difficult to realize.
        Therefore, NEST uniformly distributes the neurons of each population across the available processors to balance the compute load (\cref{sec:note_distribution}).
        The neurons are connected via synapses, which are assigned specific \textit{weights} and \textit{delays} reflecting conduction times.
        Synapse models are stored and updated on the same compute nodes that hold their postsynaptic partner neurons.
        Maintaining the complete network connectivity in computer memory enables the use of plasticity mechanisms that can modify synaptic strengths at runtime (for functional synaptic plasticity, see \cref{sec:functional-plasticity}).
        The alternative \emph{procedural connectivity} approach generates the required routing information on the fly and thereby requires fewer memory resources \citep{Roth1997,Knight2021_136}.

        Establishing synapses in a computational network model requires defining which neurons are connected.
        For specific data-driven models, the network structure can be loaded from a file, but simulators also provide built-in routines for generating connectivity.
        These routines \refnest{link:conn} range from a primitive that just connects individual source and target neurons, to high-level connection rules acting on the neuron population level \citep{Senk2022}.
        For example, the deterministic rule \textit{all-to-all} connects each neuron of a source population to each neuron of a target population.
        Probabilistic rules account for the often statistically described sparse connectivity in biological neuronal networks.
        \textit{Random, fixed in-degree} connectivity, for instance, specifies only the number of incoming connections per neuron but not which individual ones are selected as sources.
        If the connectivity is described as \textit{pairwise Bernoulli}, each pair of neurons is connected with a given probability.
        The fixed in-degree rule needs to be combined with the specification of whether multapses are allowed, whereas the pairwise Bernoulli rule excludes them by definition as each pair of neurons is considered only once.
        High-level connection rules enable efficient low-level implementations such as parallelization of the network construction.

        Pseudo-random number generators (pRNGs) are used for drawing connections according to a probabilistic rule and optionally also for setting neuron and synapse parameters (see \cref{sec:heterogeneity}). The resulting network realization will be identical if the same sequence of random numbers is sampled; this is achieved by fixing the pRNG seed.
        Random distributions sometimes have to be constrained in order to restrict the sign of a weight, e.g, according to Dale's law  \citep{Strata99_349}, or to enforce connection delays to be larger than the simulation time step, for which a typical value is $0.1~\textrm{ms}$. A longer minimum delay, for instance, $1~\textrm{ms}$, can furthermore be used to limit the necessary frequency of communication between virtual processes.

        Large-scale neuronal network models require high-performance computing.
        Employing several compute nodes in parallel not only distributes the workload but also gives access to sufficient memory for storing the network connectivity.
        Storing a single synaptic weight costs $8$ bytes in NEST \citep{kunkel2014spiking} as it sums up the effects of a set of vesicles that may differ in size, as well as of potentially different amounts of receptors. Moreover, a typical neuron in the mammalian brain has on the order of $10^4$ synapses \citep{kandel2012principles}.
        This leads to a substantial amount of resources required for large models.
        Networks with reduced neuron and synapse numbers can preserve some characteristics (e.g., firing rates) of full-scale networks if the downscaling is compensated for with informed parameter adjustments \citep{Albada15}.
        The pairwise correlation structure of the neuronal activity, however, cannot be preserved simultaneously, rendering neuroscientific simulations at natural density a necessity where correlation structure is relevant. This may be the case, for instance,
        to ensure the correct network state: correlation changes may even shift a network between linearly stable and unstable regimes.
        Data structures that keep the memory usage per MPI process constant regardless of the total number of MPI processes used in the simulation \citep{Jordan18_2} provide a potential solution, paving the way toward brain-size networks with realistic connectivity.

\section{Synaptic transmission} \label{sec:synapses}

    \subsection{From empirical data to mathematical models} \label{sec:synaptic_transmission_from_data_to_model}
        Electrochemical signaling between neurons is mediated by various receptor types, expressed post- and presynaptically.
        Different receptor types trigger different physiological responses. Ligand- or voltage-gated ion channels influence ionic flows through the membrane, with distinctive kinetics for each receptor type (examples include AMPA and NMDA). Metabotropic receptors trigger intracellular biochemical signaling cascades with downstream actions that are typically not instantaneously noticeable but mediate physiological adaptation processes. \emph{Electrical synapses} (\emph{gap junctions}) are transmembrane channels that form a direct electrical and biochemical coupling between the cytosol of two adjacent cells. Compared to \emph{chemical synapses}, they provide increased speed as the signal does not need to be converted from electrical to chemical and back across a synaptic cleft.
        The composition of receptor types on a neuron's synapses, their spatial distribution on the dendritic tree and cell body, and their individual, instantaneous efficacy and response kinetics determine how and at which timescales the neuron filters and integrates its many presynaptic inputs. We will focus here on chemical synapses.

        The amplitude of the postsynaptic response is proportional to the synapse's strength or \emph{weight}, which depends on the amount and types of both neurotransmitters and receptors as well as the state of the postsynaptic neuron. Synaptic weights can be estimated from paired recordings, usually \emph{in vitro}, to avoid background activity that confounds the measurements. However, here it should be taken into account that the synaptic weight obtained from paired-cell recordings is determined by a combination of biophysical properties, e.g., postsynaptic receptor density, amount of released neurotransmitters, reuptake kinetics, or existence of more than one connection between a pair of cells (multapses). Hence, the terms \emph{strength} and \emph{weight} refer to an effective, phenomenological quantity. Mapping the corresponding parameters to the \emph{in vivo} condition is nontrivial because experimental conditions like temperature and extracellular fluid may differ, as well as a high-conductance network state affecting the measured quantities such as time constants \citep{Destexhe2003high, maksimov2018criteria}. Change of synaptic strengths over time is discussed in \cref{sec:functional-plasticity}.

        When a presynaptic neuron has emitted an action potential, the signal that arrives at the postsynaptic neuron can be observed as a postsynaptic potential (PSP): the deflection in the somatic membrane voltage caused by the incoming spike. Alternatively, synaptic currents (PSCs) may be measured at different holding potentials using voltage clamp recordings. From a PSC or PSP, the weight of the synapse can be derived. Synapses are often modeled as injecting a current into the postsynaptic cell or acting as a conductance, the current of which is proportional to the difference between the membrane potential and a synapse- or receptor-specific reversal potential. In the simplest approximation of the postsynaptic response kinetics, the time course of this current or conductance may be modeled as a Dirac delta function causing a step increase in the membrane potential or current, respectively. With increasing levels of complexity, the time course of the PSC (or postsynaptic conductance, PSG) can be approximated by an instantaneous rise followed by an exponential decay or by a double exponential with separate time constants for the rising and the decaying phase \citep{GerKisNauPan2014}.

        Beside spatially precise communication via synapses, the spatially more diffuse process of \emph{neuromodulation} can alter the excitability of neurons and affect synaptic plasticity (see \cref{sec:functional-plasticity}).
        Neuromodulation is achieved by the release of a neurotransmitter with less detail in the connectivity patterns than in typical synaptic (for instance, glutamatergic) neurotransmission. The neuromodulator, such as dopamine or serotonin, is typically released from a neuron whose cell body lies in a small, circumscribed nucleus in the brain but which projects broadly and affects many downstream targets simultaneously. The precise spatiotemporal profile of neurotransmitter concentration is often approximated in models by assuming the neuromodulator diffuses through extracellular space, referred to as \emph{volume transmission}. Simulating the diffusion process entails solving the Laplace equation\textemdash{}an equation that involves the spatial gradient and divergence operators, requiring a different type of solver than those that solve the neuronal network system dynamics. Instead of a detailed representation of the geometry of extracellular space  (for instance, based on the finite-element method), the medium may be assumed to be spatially homogeneous, and diffusion can even be assumed to occur instantaneously, considerably simplifying the model and its computational requirements \citep{potjans2010enabling}.

        Neurons have a spatial extent, and their dendrites often exhibit intricate branching patterns. Consequently, the spatial collocation of synapses on the dendrites has important consequences for the neuron's response to input.
        Dendritic responses are often nonlinear, as dendrites are studded with a high density of voltage-gated channels, which, combined with intracellular responses like calcium signaling, can cause a nonlinear interaction between nearby synaptic inputs. In addition, the dendrite itself can exhibit action potentials distinct from a somatic action potential, for instance, involving a local, intracellular calcium transient \citep[see, e.g., ][]{Larkum2022}. The (local) change in membrane potential and conductance, in turn, affects the integration at adjacent synapses in the branch.
        The triggering of dendritic action potentials by co-activated and co-located synapses and their effects on the somatic dynamics can be accounted for in simple point neuron models by including nonlinearities in synaptic input currents \citep{Jahnke12_041016,Bouhadjar2022sequence}.
        For a more fine-grained analysis, multicompartment models are commonly used. In these models, each neuron consists of dozens or hundreds of compartments, each equipped with a distinct type of dynamics and parameterization and coupled to neighboring compartments according to Ohm's law \citep{GerKisNauPan2014}.
        Multicompartment models permit integrating experimental data at a highly detailed level of description but are computationally and conceptually much more complex and demanding. On the other hand, some biophysical details like synaptic adaptation can be adequately modeled without the need to address the microscopic biophysics of synaptic vesicles but can be treated phenomenologically by adding one or a few extra continuous state variables to the model \citep[e.g.,][]{tsodyks1998neural,brette2005adaptive}.

    \subsection{From mathematical models to simulation}

        NEST integrates equations for neurons with linear subthreshold dynamics exactly \citep{Rotter99a} and uses standard numerical solvers for nonlinear neuron models. For synapses, an efficient approach is to specify the characteristic time evolution of some postsynaptic quantity, such as current or conductance, as a linear system of equations. If responses sum linearly across a neuron's synapses, they can be lumped together into a single or a few state variables and do not have to be stored and updated for each synapse separately. For this reason, the postsynaptic response is typically specified as part of the (postsynaptic) neuron model. As described in \cref{sec:synaptic_transmission_from_data_to_model}, the postsynaptic kernel could be, for example, a Dirac delta function (causing an instantaneous jump in the postsynaptic membrane potential), an instantaneous rise followed by an exponential decay, or a double-exponential function with a finite rise time. Furthermore, because they are linear, solving these equations does not require a numerical solver but only multiplication with a constant at each time step \citep{morrison2007exact}. The reduction to a simple multiplication generally makes the solution much more precise: the computed values are closer to the mathematically ``true'' solution and more efficient to compute. Thus, simulations of networks with many synapses become feasible. Multiple types of synapses can be easily incorporated into this scheme by grouping them according to their kinetics, for instance, into a separate AMPA and NMDA group \refnest{link:multisyn-nestml-model}.

        In simulations of large networks, the layout of data structures in memory and communication can become bottlenecks. Conceptual modeling decisions can interact with data layouts; for example, the synaptic delay can be chosen as a property of the synapses or the pre- or postsynaptic neurons. In the point-neuron framework, the delay is assigned to either a neuron's axonal or dendritic side, implying different biophysical interpretations and simulation outcomes. The biophysical object of a synapse is not necessarily represented in code by a specific software object but distributed into a presynaptic and a postsynaptic component. In the instantiation of a particular model, these components may not even live on the same compute node. As noted in \cref{subsec:conn2}, NEST stores synapses on the process containing the postsynaptic neuron.

        In simulations using parallel computing, spike events and potentially other quantities such as synaptic weights have to be communicated between threads, processes, or across a computer network (\cref{sec:note_scalability}). Parallel computing presents a set of unique design requirements because the evolution of the dynamical model needs to occur synchronously, lest the model's state becomes internally inconsistent when some parts of it have become desynchronized in time. This requirement can be addressed by instituting a minimum, nonzero transmission delay for each synaptic connection in the model. A delay between the presynaptic spike and the resulting postsynaptic response effectively decouples neurons for this time window so that events can be transmitted across the computer network in a regular cadence at the end of each window (\cref{sec:note_precise-spikes}). This decoupling allows simulations to scale to many compute nodes \citep{Morrison08_267}.

        From a mathematical modeling point of view, gap junctions are much simpler than chemical synapses; their delay is negligible, and they do not filter the input. However, modeling gap junctions numerically can be challenging because they entail an instantaneous coupling between compartments. The \emph{waveform relaxation} technique helps retain simulation efficiency when combining gap junctions with a numerical simulation method that takes advantage of a minimum, nonzero synaptic delay. Each neuron is considered a separate subsystem in this technique, and the gap junction coupling terms (current flowing from one neuron into the other) are solved iteratively. This solution requires exchanging data (in particular, membrane potentials) between the gap-junction coupled neurons only at the end of each minimum delay step, thus limiting the necessary communication frequency. In exchange, it requires only a modest increase in computation and the size of the communicated packets since solving the forward dynamics of each separately considered neuron needs to be repeated only once per iteration of the waveform relaxation algorithm \citep{hahne2015unified, jordan2020efficient}.

\section{Structural Plasticity} \label{sec:structural-plasticity}
    The models introduced earlier in this chapter had static connectivity (see \cref{sec:connectivity}).
    However, macroscopic observations of the brain have revealed that the connections in cortical networks continually change as new synapses form and others dwindle and disappear \cite[for a review, see][]{stettler2006axons}.
    This rewiring is a lifelong process to encode experiences but happens extensively during development and recovery from lesions in the brain tissue \cite[for a review, see][]{butz2009activity}.
    The underlying mechanisms introducing dynamics into the connectivity are summarized as \emph{structural synaptic plasticity}.

    \subsection{From empirical data to mathematical models}
        Including structural plasticity mechanisms into a synapse model can increase its biological plausibility, e.g., regarding learning, development, reformation after lesions \citep{butz2013simple} or topographic map formation \citep{bamford2010synaptic}. Structural plasticity might also be the basis for associative connections \citep{gallinaro2018associative} and metaplasticity \citep{kalantzis2009structural}. However, plasticity mechanisms capable of generating network connectivity in a principled fashion can also be helpful in other ways. First, they can help reduce dependence on cumbersome and expensive connectivity recordings in animals (see \cref{sec:connectivity}). Second, they can serve a range of functional purposes. For example, they can enhance learning performance \citep{bellec2017deep} or increase the storage efficiency of long-term memories and, by that, prevent catastrophic forgetting \citep{knoblauch2017impact}. Plasticity mechanisms also frequently serve a homeostatic function. In general, the term ``homeostasis'' refers to a range of vital physiological processes that assist organisms in maintaining internal states (such as body temperature, blood sugar levels, and heart rate) at optimal levels. Likewise, homeostatic plasticity maintains quantities such as spiking activity or numbers of connections at an energetically or computationally favorable set-point \citep{Turrigiano2012}. Efficient pruning of the connectivity and preserving sparse connectivity \citep{kappel2015network} can help save energy and optimize the usage of limited synaptic resources, which is particularly important in neuromorphic computing \citep{bellec2017deep,billaudelle2021structural,george2017activity}.

        To understand the process of developing a comprehensive and accurate mathematical model of structural plasticity, the following paragraphs sketch the steps involved in creating the model suggested by \citet{butz2013simple} as an example.
        This model is based on the observation that the creation and deletion of synapses can bring the postsynaptic neuron's firing rate into a certain physiological range.
        The authors consider synaptic elements, namely axonal boutons on the presynaptic side and dendritic spines on the postsynaptic side.
        When two such elements are combined, a synapse is created.
        The dynamics of the number of synaptic elements for each neuron depends on a readiness variable $c$ (associated with the calcium concentration), which indicates the propensity of a neuron to grow synapses.
        The readiness variable is a low-pass filtered version of the spiking activity and thus approximates the neuron's instantaneous rate up to a scalar multiplier.

        The algorithm comprises four steps, which are repeated until the connectivity converges.
        First, it continuously updates the spiking activity of the neurons since each neuron's mean firing rate influences the creation of synaptic elements.
        Second, it updates the readiness $c$ for each neuron:
        \begin{equation}
            \frac{\mathrm{d}c}{\mathrm{d}t}=-\frac{c(t)}{\tau}+ \beta \delta\left(t-t_j^\text{f}\right)\;,
            \label{eq:readiness}
        \end{equation}
        i.e., $c$ decays exponentially with the time constant $\tau$ and increases by a fixed amount $\beta$ whenever the neuron $j$ spikes at $t_j^\text{f}$, where $\delta(\cdot)$ denotes the Dirac delta function and $\text{f}$ stands for ``firing''.
        Third, a homeostatic rule drives the neuron to reach and maintain a target activity by deleting postsynaptic elements if the instantaneous activity is higher than the target activity or creating synaptic elements if the current activity is lower than the target activity.
        A growth curve defines the speed of these modifications towards a target calcium concentration $c_\text{target}$ by means of a growth rate $\nu$ and can be expressed as a linear function
        \begin{equation}
            \frac{\mathrm{d}z}{\mathrm{d}t}=\nu\left(1-\frac{c(t)}{c_{\text{target}}}\right)\;.
            \label{eq:linear}
        \end{equation}
        Alternatively, a downward shifted Gaussian or other more complex function may be used, as long as it has a zero-crossing with a negative gradient that allows convergence.
        If the value of $z$ increases or decreases by 1, the neuron grows or deletes a synaptic element, respectively.
        The algorithm creates new connections between randomly chosen synaptic elements from the available set in the fourth and last step.
        This set comprises synaptic elements generated in previous iterations that are not yet connected and the connection partners of deleted synaptic elements.

        Beyond this specific example, there exists a range of different structural plasticity models: some have rules for deleting, some for forming synapses, and some for both \cite[for a book, see][]{vanooyen2017rewiring}.
        Often the algorithm prunes synapses that do not have the chance to become active again \citep{iglesias2005dynamics}, are the weakest according to a specific metric \citep{hawkins2016neurons, roy2014liquid}, or have too little causal correlation between pre- and postsynaptic spikes \citep{bourjaily2011synaptic}. Sometimes the mechanism's objective is to maintain a preset number of connections \citep{bellec2017deep}, preserve short-range connections \citep{butz2014homeostatic}, or prune until the connectivity has converged to the most efficient constellation \citep{iglesias2005dynamics}.
        In many algorithms, the determining factors for rewiring are the pre- and postsynaptic activity and the vicinity of other synapses.
        In general, one can divide structural plasticity mechanisms into two categories: Hebbian structural plasticity, which leads to an increase in the number of synapses during phases of high neuronal activity and, conversely, a decrease in phases of low neuronal activity; and homeostatic structural plasticity, which balances these changes by removing and adding synapses \citep{fauth2016opposing}.

    \subsection{From mathematical models to simulation}
        The NEST implementation of the discussed particular structural plasticity mechanism \citep{diaz2016automatic} updates the network connectivity at time intervals that are long compared to the computational time step used to update the neurons, based on experimental observations \refnest{link:struc-plast}.
        This slow timescale makes the algorithm more efficient, as the available synaptic elements do not need to be calculated and communicated at every time step.
        However, when using structural plasticity to generate connections in a network, note that convergence is not guaranteed but determined by the growth rate, network connectivity, and network activity; thus, visual guidance is advised (see \citealp{nowke2018toward} and \cref{sec:note_convergence}).

\section{Functional Plasticity} \label{sec:functional-plasticity}

    The strength of a synapse is usually parameterized by a single static value, the synaptic efficacy or synaptic weight (see \cref{sec:synapses}).  However, existing synapses can grow stronger or weaker as an effect of a variety of biophysical mechanisms on both the pre- and postsynaptic side, phenomena collectively known as \emph{functional synaptic plasticity}. These adjustments of synaptic efficacies are likely to form the basis of learning and memory processes in the brain. Thus, this section addresses the temporal evolution of synaptic efficacies and the underlying mechanisms.

    \subsection{From empirical data to mathematical models}

        Over 70 years ago, Hebb famously postulated that neurons that fire together wire together \citep{hebb1949organisation}. Since then, many phenomenological models of functional plasticity have been derived and developed. Introducing categories brings some order into the vast landscape of models, even if they do not have clear-cut boundaries and often overlap. Four categorizations are common. First, with respect to the timescale: while short-term plasticity models cover timescales from milliseconds to seconds, long-term plasticity models cover minutes to hours, and homeostatic plasticity models (e.g., synaptic scaling) even up to days \citep{magee2020synaptic,morrison2008phenomenological}. While structural plasticity can occur over the course of hours \citep{okabe1999continual}, timescales of functional plasticity are typically shorter. One needs structural plasticity to create new synapses (happening on a long timescale) which then grow stronger via functional plasticity (happening on a faster timescale). Vice versa, synapses that have grown weak are more likely to be pruned.
        To date, models usually include either functional or structural plasticity but not both together, and the effects of these mechanisms on synaptic learning are thus studied independently.
        A second categorization distinguishes functional plasticity according to the mechanisms involved: Unsupervised learning rules are based on unlabeled data, supervised learning rules involve a target signal, and reinforcement learning rules function via rewards \citep{magee2020synaptic,morrison2008phenomenological}.
        A third categorization considers the number of factors that constitute the update formula of the synaptic efficacy: Standard correlation-based rules usually involve two factors, the pre- and postsynaptic spiking, as opposed to three-factor models that involve an additional modulatory signal, e.g., neuromodulation.
        Fourth, based on activity dependence, plasticity mechanisms occur in two counteracting forms: in Hebbian-type mechanisms, higher activity levels of the pre- and post-synaptic neurons lead to strengthening of the (positive or negative) weights, whereas in homeostatic mechanisms, to their weakening \citep{fauth2016opposing}.
        Without further constraints, Hebbian-type plasticity may lead to a positive feedback loop and, consequently, substantial changes in synaptic weights and network activity. In contrast, homeostatic synaptic plasticity pushes the synaptic efficacy up if activities are low and down if neuronal activities are high, inducing a negative feedback loop and stabilizing the dynamics.

        Experimental studies show that the efficacy of a synapse can change for a short time window of hundreds to thousands of milliseconds depending on the history of the presynaptic spikes \citep{Tsodyks97,markram1998differential,gupta2000organizing}.
        This phenomenon is termed \emph{short-term plasticity (STP)}, or more precisely, \emph{short-term facilitation (STF)} if the efficacy is elevated, and \emph{short-term depression (STD)} if the efficacy is decreased.
        The biophysical mechanism underlying STP is the dynamics of vesicle pools and spike-triggered exocytosis.
        On the one hand, after the generation of a spike, calcium accumulates in the presynaptic axon terminal, increasing the probability of neurotransmitter release, which enhances the synaptic efficacy and thus causes STF. On the other hand, repetitive firing leads to the depletion of vesicles and saturation of postsynaptic receptors, which decreases the efficacy and thus causes STD.
        The mechanisms for STF and STD are counteracting, and a combination of both can be present in the same synapse. Depending on the synapse or neuron type, one of them may be more pronounced.
        These phenomena form the basis for many STP models \citep[for a review, see][]{Zucker2002}. The following paragraph outlines one possible modeling approach by using the example of the model proposed by \citet{Tsodyks97}.

        The starting point is the view introduced in \cref{sec:connectivity,sec:synapses}: a neuron $k$ receives spikes from neuron $j$ over a synapse with the weight $w_{jk}$. Now, to make the static synaptic weight a dynamical variable, $w_{jk}$ is multiplied by a time-dependent scaling factor $f_\text{a}(t)$, modeled by a set of three coupled differential equations:
        \begin{align}
            \frac{\text{d}f_\text{a}}{\text{d}t} &= - \frac{f_\text{a}}{\tau_\text{i}} + u^+ f_\text{r}^- \delta\left(t-t_j^\text{f}\right),
            \label{eq:fa}\\
            \frac{\text{d}f_\text{r}}{\text{d}t} &= \frac{f_\text{i}}{\tau_\text{r}} - u^+ f_\text{r}^- \delta\left(t-t_j^\text{f}\right),\,\text{and}
            \label{eq:fr}\\
            \frac{\text{d}f_\text{i}}{\text{d}t} &= \frac{f_\text{a}}{\tau_\text{i}} - \frac{f_\text{i}}{\tau_\text{r}},
            \label{eq:fi}
        \end{align}
        describing the utilization of synaptic resources by each presynaptic spike arriving at time $t_j^\text{f}$.
        Here, $f_\text{a}(t)$, $f_\text{r}(t)$, and $f_\text{i}(t)=1-f_\text{r}(t)-f_\text{a}(t)$ denote the fractions of active, recovered, and inactive synaptic resources, respectively, and $\delta(\cdot)$ the Dirac delta function.
        The superscripts ``$-$'' and ``$+$'' refer to the values of the associated variables before and after their update.
        Each presynaptic spike increases the active synaptic resources, i.e., the synaptic weight, and simultaneously reduces the available (recovered) resources by an amount proportional to $u^+$.
        The utilization variable $u^+$ represents the probability of vesicle release, controlled by the calcium concentration in the axon terminal. In the absence of STF, this utilization is constant. However, in facilitating synapses, it is dynamic and evolves according to
        \begin{equation}
            \frac{\text{d}u}{\text{d}t} = - \frac{u}{\tau_\text{fac}} + U\left(1-u^-\right)\delta\left(t-t_j^\text{f}\right),
        \end{equation}
        where the parameter $U$ corresponds to the amplitude of the postsynaptic current (the synaptic weight) in response to a single isolated presynaptic spike.
        The time constants $\tau_\text{r}$, $\tau_i$, and $\tau_\text{fac}$ describe the recovery time from synaptic depression, the decay of the postsynaptic currents, and the decay of the utilization, respectively.
        Despite its simplicity, this phenomenological model approximates experimental findings well \citep{Tsodyks97}.

\begin{figure}[t!]
            \centering
            \includegraphics[width=\linewidth]{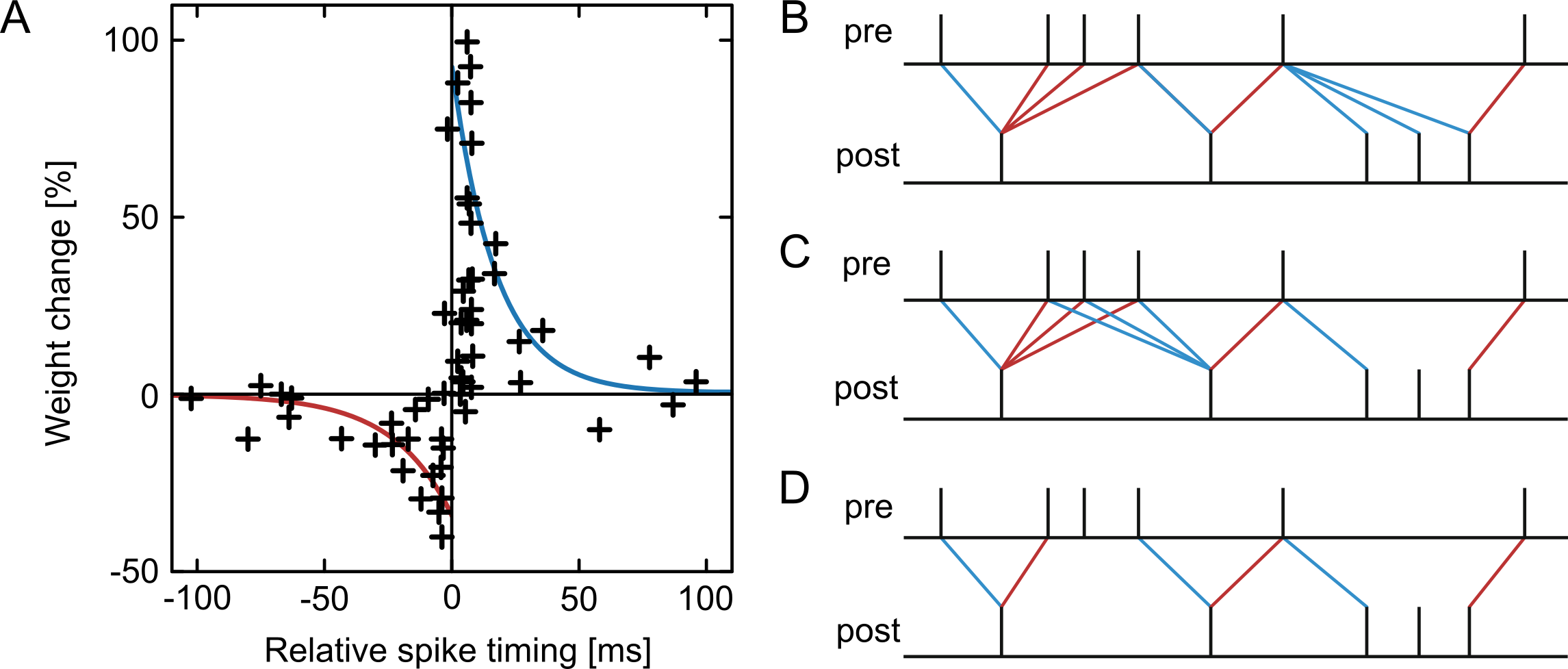}
            \caption{
            {\bf Spike-timing-dependent plasticity (STDP).}
            {\bf A} Weight change expressed as a function of relative pre- and postsynaptic spike timing for an example of STDP with an anti-symmetric (and slightly asymmetric), Hebbian learning window. Markers correspond to empirical data from \citet{bi1998synaptic}. Solid lines show exponential approximations used in the model (red indicates depression and blue indicates potentiation in all panels). An anti-Hebbian window would look similar but mirrored about the vertical axis.
            {\bf B-D} Three nearest-neighbor spike pairing rule variants for STDP.
            {\bf B} \emph{Symmetric:} each presynaptic spike is paired with the last postsynaptic spike, and each postsynaptic spike is paired with the last presynaptic spike.
            {\bf C} \emph{Presynaptic centered:} each presynaptic spike is paired with the last postsynaptic spike and the next postsynaptic spike.
            {\bf D} \emph{Reduced symmetric:} as in panel C, but only for closest pairs.
            Adapted from Fig.~7 in \citet{morrison2008phenomenological}.
            }
            \label{fig:stdp}
        \end{figure}

        The modifications of the synaptic efficacy by STP occur only during presynaptic firing and last for a few hundred milliseconds.
        After presynaptic firing has stopped, the synaptic resource variables, and hence the synaptic weight, return to their resting states $f_\text{a}=0$, $f_\text{r}=1$, $f_\text{i}=0$, $u=0$, and $w_{jk}$.
        In contrast, \emph{spike-timing-dependent plasticity (STDP)} has a prolonged effect on synaptic efficacy and thus constitutes a form of long-term plasticity.
        This form of plasticity was discovered in several spike pairing experiments where a pre- and a postsynaptic neuron were repetitively stimulated to emit spikes at a predefined interval \citep{markram1997physiology, bi1998synaptic}.
        Reviews of the experimental findings can be found in \citet{caporale2008spike,markram2012spike} and \citet{brzosko2019neuromodulation}.
        Although the results of these studies vary across cell types and pairing protocols, they all find that the induced change of the synaptic efficacy depends on the precise time difference $\Delta t = t_\text{post} - t_\text{pre}$ between a pair of pre- and postsynaptic spikes.
        The mapping between the sign and value of the weight changes and the time lags between pre- and post-synaptic spikes is described by the \emph{STDP kernel}, which can take either a Hebbian or an anti-Hebbian form and is sometimes modeled as either symmetric or anti-symmetric \citep[all forms illustrated by Fig.~2 in]{li2014activity}. In nature, however, the size and shape of the potentiation and depression windows might differ, leading to an overall asymmetric window (see \cref{fig:stdp}A).
        Generally, a postsynaptic spike occurring slightly after the presynaptic spike ($\Delta t > 0$) induces \emph{long-term potentiation (LTP)}, whereas a postsynaptic spike occurring slightly before the presynaptic spike ($\Delta t < 0$) induces \emph{long-term depression (LTD)}.
        Thus, STDP can encode a causal relationship between the firing of the pre- and postsynaptic neuron and the synaptic weight change.
        Since this finding follows Hebb's principle, this type of STDP belongs to the class of Hebbian plasticity rules.

        Formalizing this robust finding based on the above experimental data allows for mathematical treatment.
        \citet{morrison2008phenomenological} developed a phenomenological model with only a few free parameters, which reproduces the experimental observations without referencing the underlying molecular mechanisms.
        They restricted the observables that can enter the plasticity rule to locally available ones because biologically plausible phenomenological models should only contain terms that are identifiable with mechanisms that exist in biology.
        In the case of STDP, the synaptic weight change depends on the pre- and postsynaptic spike times and potentially also on the current synaptic weight.
        Experiments demonstrating that action potentials back-propagating through the dendritic tree convey information about a postsynaptic spike support the fact that postsynaptic spikes can be available at the synapse \citep{markram1997physiology}.

        The dependence of LTP and LTD on $\Delta t$ is usually captured by exponential functions with decay times $\tau_{\pm}$.
        \citet{morrison2008phenomenological} give a simple model of the weight change $\Delta w$ for pair-based STDP:
        \begin{equation}
        \begin{aligned}
        \Delta w_{+} &= F_{+}(w)\, e^{-\frac{|\Delta t|}{\tau_{+}}} &\text{if}\,\Delta t > 0, \\
        \Delta w_{-} &= -F_{-}(w)\,e^{-\frac{|\Delta t|}{\tau_{-}}} &\text{if}\,\Delta t \leq 0,
        \end{aligned}
        \label{eq:stdp}
        \end{equation}
        where the functions $F_{\pm}$ capture the dependence on the current weight $w$ and have to be specified further by fitting them to experimental data \citep{morrison2007spike,vanrossum2000stable} (\cref{fig:stdp}A).
        A spike pair can be defined in different ways: for example, each presynaptic spike can be paired with the most recent preceding postsynaptic spike and vice versa, which is one variant of the class of  \emph{nearest neighbor schemes} (see \cref{fig:stdp}B--D), whereas in the \emph{all-to-all scheme}, each presynaptic spike is paired with all preceding postsynaptic spikes \citep{morrison2008phenomenological,burkitt2004spike}.

        The STDP model described above can serve as a starting point for designing extended models to describe more nuanced experimental results, for example by including the postsynaptic membrane potential as an additional modulatory factor beyond the pre- and postsynaptic spikes. Along these lines, \citet{clopath2010voltage} and \citet{clopath2010connectivity} account for the effects of voltage-dependent receptors and channels. Their approach is based, among other findings, on experiments showing that the same spike pairing protocol can, depending on the postsynaptic membrane potential $V_\text{m}$, induce no change in synaptic weights at all, LTD, or LTP \citep{Ngezahayo2000synaptic}. While a $V_\text{m}$ smaller than an experimentally determined threshold potential $\Theta_{-}$ induces neither LTD nor LTP, an intermediate membrane potential $\Theta_{-} < V_\text{m} < \Theta_{+}$ triggers LTD, and a high $V_\text{m} > \Theta_{+}$ enables LTP. To capture this behavior, the mathematical description of the plasticity rule contains terms for facilitation and depression that are active based on these conditions of the membrane voltage, formally expressed as Heaviside step functions. With this mechanism, \citet{clopath2010voltage} were able to reproduce the complex frequency dependence of the synaptic weight changes in spike pairing experiments \citep{sjostrom2001rate}.

        In \citet{Urbanczik2014}, the postsynaptic membrane potential is included as a modulating factor. This plasticity rule, in particular, applies to synapses that connect to the dendrite of a postsynaptic neuron. Experiments show that presynaptic spikes that do not cause postsynaptic spikes lead to a depression of synaptic weights whose strength increases with increasing dendritic voltage \citep{Artola1990different}. From this observation, \citet{Urbanczik2014} conclude that the synaptic weights are adjusted such that the dendritic voltage assumes high values if and only if the soma of the postsynaptic neuron emits spikes. Therefore, in this rule, the difference between the dendritic voltage and the somatic activity drives the synaptic weight change.

        Instead of the postsynaptic membrane potential, a third factor could also be a neuromodulator concentration, which is motivated by experimental studies \citep[for a review, see][]{pawlak2010timing} and by the fact that they provide a biologically plausible implementation of reward signals \citep{worgotter2005temporal}.

    \subsection{From mathematical models to simulation}
        The STP implementation in NEST \refnest{link:tsodyks-depress,link:tsodyks-fac} exploits several practical properties of the corresponding differential equations for their numerical integration.
        Since the synaptic resources are conserved (i.e., the fractions $f_\text{r}$, $f_\text{a}$, and $f_\text{i}$ add up to 1), $f_\text{a}$ can be eliminated from the system.
        Furthermore, thanks to its linear form, the system of coupled differential equations \cref{eq:fr,eq:fa,eq:fi} can be integrated exactly between two consecutive presynaptic spikes \citep{Rotter99a}.
        Concretely, the joint state of $u$, $f_\text{r}$, and $f_\text{a}$ can be iteratively evolved by multiplying the state at the previous presynaptic spike with a \emph{propagator matrix}.

        To simulate STDP, \cref{eq:stdp} needs to be calculated efficiently \citep{morrison2008phenomenological}.
        Having restricted the model parameters to those locally available at the synapse facilitates the implementation in software \refnest{link:stdp}.
        These constraints also improve the model's performance, since network simulators running on distributed systems take advantage of a limited need for global access to variables to reduce memory consumption and high-latency communication between compute nodes \citep{stapmanns2021event,morrison2005advancing}.
        The all-to-all pairing scheme can be efficiently implemented using a specific update scheme of the synaptic traces.
        These traces represent a fading memory of past spikes at the synapse without explicit knowledge of all past spike times \citep{morrison2007spike,morrison2008phenomenological}.
        If a pre- or postsynaptic spike occurs, the corresponding trace and synaptic weight are updated, while no actions need to be performed in the periods in between.
        Defining the exact order of updates in a plasticity model, particularly with regard to pre- and postsynaptic spike timing, indicated by the ``+'' and ``-'' in \cref{eq:fa,eq:fr,eq:fi} is crucial and
        facilitated by the high-level language NESTML \refnest{link:weight-norm}.

        More complex learning scenarios, like reinforcement learning, are made possible by advanced plasticity rules, which, for example, depend on the postsynaptic membrane potential or neuromodulators \citep{weidel2021unsupervised}. However, these rules typically make it more difficult to discover an effective implementation.
        For example, neuromodulators (e.g., dopaminergic signals) affect several nearby synapses through \emph{volume transmission} requiring a notion of physical 3D space (see \cref{sec:synaptic_transmission_from_data_to_model}).
        Moreover, the presence of time-continuous signals in some of these advanced plasticity rules necessitates the storage of the signal history if one wishes to keep the efficient \emph{event-driven} scheme of updating the synaptic weights only at presynaptic spike times \citep{stapmanns2021event}.
        Depending on how the data structures are laid out in a simulator, accessing the continuous third-factor variables can be difficult or computationally costly because they need to be queried for every spike at every synapse. Generally, rules that require only spike times are more efficient in memory and compute time than rules that depend on the entire history of variables, like the membrane potential trace.

        Given access to the synaptic weights, another form of functional plasticity, \emph{weight normalization}, can be realized. It entails keeping the total sum, or a norm of all incoming synaptic strengths of a neuron constant by re-normalizing all its synaptic weights \refnest{link:weight-normalization}. Since, in the brain, these weight changes happen on timescales of several hundreds of milliseconds, the iterative re-normalization takes place on a coarse time grid, increasing the operation's efficiency.

        Other advanced plasticity rules include, for example, a third-factor postsynaptic dendritic current \citep{Urbanczik2014} or inhibitory plasticity  \citep{vogels2011inhibitory}. The discovery of new plasticity rules can be, to a certain degree, even automated \citep{jordan2021evolving}. Furthermore, state models with \emph{synaptic tagging and capture} (STC), described, e.g., in \citet{Barrett2009}, incorporate even plasticity effects beyond synapse-specific ones. Ultimately, state-of-the-art computational plasticity models transcend the simple STP and STDP models \cite[see, e.g.,][]{Mongillo2008}. Algorithmically, however, these complicated models often use a combination of plasticity mechanisms and thus can be synthesized from such a base stack of simpler models.

\section{Heterogeneity} \label{sec:heterogeneity}
    Complexity and heterogeneity are ubiquitous and well-established design principles in neurobiological systems \citep{Koch1999}, covering a multitude of components and mechanisms at various spatial and temporal scales. From an information processing perspective, such variability is a fundamental component of the system, as it determines the types of computations a given circuit can perform and constrains the representational expressivity of its dynamics \citep{duarte2019leveraging}.

    \subsection{From empirical data to mathematical models} \label{sec:heterogeneity_from_experiments_to_models}
        Biological synaptic connectivity is highly diverse in most of its constituent properties, including the type of neurotransmitter used, the composition of presynaptic vesicles and docking proteins (affecting release probability), the postsynaptic receptor composition (affecting efficacy and kinetics of the elicited response), transmitter re-uptake and re-use, and the involvement of gliotransmission \citep[see, e.g.,][]{Parpura2010gliotransmission}, but also properties characterizing signal propagation such as axon diameter and conductance velocity \citep{Girard01,Liewald14_541,Muller18_255}. These various types of diversity translate to a high degree of heterogeneity in phenomenological parameters characterizing mathematical models of the synaptic connectivity and dynamics, such as the synaptic weight \citep{Song05_0507,Lefort09_301,Koulakov09_3685,Avermann12,Ikegaya13_293}, synaptic time constants \citep{Kuhn04,Roxin11_16217}, response latencies (synaptic delays; \citealp{Brunel99,Roxin11_16217}), and parameters specifying the plasticity dynamics \citep{Kampa07_456}.
        In addition, biological neuronal networks exhibit a high degree of heterogeneity in the anatomical connectivity structure, such as the total number of inputs and outputs per neuron (in/out-degrees; \citealp{markram1997physiology,Feldmeyer99_169,Feldmeyer02_803,Feldmeyer06_583,Stepanyants07,Roxin11_8}), and the composition of presynaptic source and postsynaptic target neuron populations.

        Previous theoretical work on recurrent neuronal networks shows that heterogeneity in single-neuron properties or connectivity broadens the distribution of firing rates \citep{Vreeswijk98_1321,Roxin11_16217} and affects the stability of asynchronous or oscillatory states as well as the level of synchrony \citep{Tsodyks1993_1280,Golomb93_4810,Brunel99,neltner2000_1607,Denker04,Roxin11_8,Mejias2012_228102,Pfeil16_021023}. A large number of theoretical and experimental studies point at the benefit of heterogeneity for the information processing capabilities of neuronal networks \citep{Stocks2000,Shamir06_1951,Chelaru2008_16344,Osborne2008,Padmanabhan10_1276,Marsat2010,Holmstrom2010,Mejias2012_228102,Yim2013_032710,lengler2013_e80694,Mejias2014,duarte2019leveraging}.
        Therefore, modeling studies aiming at understanding the dynamical and functional principles of biological neuronal networks need to account for the synaptic (and other types of) heterogeneity.

\begin{figure}[t!]
            \centering
            \includegraphics[width=0.6\linewidth]{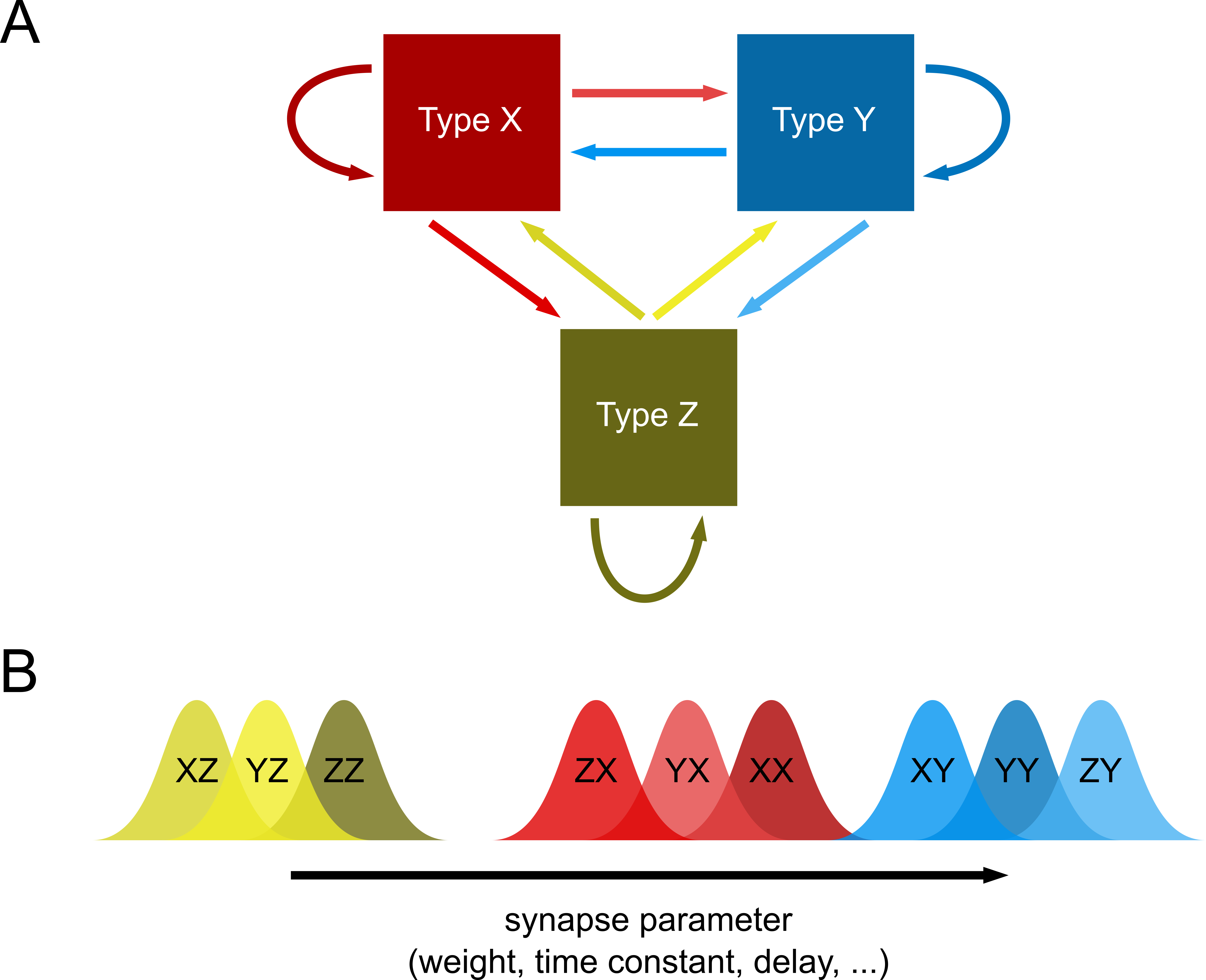}
            \caption{
            {\bf Specific and unspecific heterogeneity in synaptic connectivity.}
            {\bf A} Sketch of a neuronal network comprising three populations of neurons of type $X$, $Y$, and $Z$ (boxes). The properties of the different projections (arrows), such as the number of synapses, the synaptic weights, synaptic time constants, or synaptic delays, depend on the types of pre- and postsynaptic neurons. We refer to the resulting synapse-type specific diversity as \emph{specific heterogeneity}.
            {\bf B} For each type of projection $\{PQ\}$ from population $Q$ to population $P$ ($P,Q\in\{X,Y,Z\}$), the synaptic parameters are distributed (illustrated here with bell-shaped curves).
            We refer to this form of variability as \emph{unspecific heterogeneity}.
            The parameters characterizing each distribution, such as the mean (horizontal position of each curve) or the variance, are usually synapse-type specific.
            }
            \label{fig:synaptic_heterogeneity}
        \end{figure}

        Depending on the type of synaptic heterogeneity, its implementation in mathematical models may follow different strategies.
        Synaptic heterogeneity is expressed on local scales, such as in the connections between neurons in a given layer of a cortical column, and on large scales, such as in cortical inter-area connections.
        One form of this heterogeneity results from cell-type, layer, or area specificity.
        It reflects the anatomical and electrophysiological diversity of neurons in different brain regions and, in addition, emerges from specific interactions with other components of the nervous system or with the environment during brain development and learning.
        In mathematical models, this specificity is usually accounted for by subdividing the network into several populations representing different cell types or brain regions and applying distinct connectivity, synapse, and plasticity parameters for each pair of populations (\cref{fig:synaptic_heterogeneity}A).

        Another form of synaptic heterogeneity appears in an unspecific, quasi-random manner.
        It refers to variations in the synaptic characteristics across an ensemble of neuron pairs of seemingly identical type, for example, connections between a group of neurons with similar morphological and electrophysiological characteristics located in the same layer of a given cortical column (\cref{fig:synaptic_heterogeneity}B).
        Similarly to the cell-type-, layer-, or area-specific diversity described above, the unspecific forms of heterogeneity are partly caused by synaptic plasticity, i.e., by adapting synaptic parameters during learning and development.
        In this respect, unspecific heterogeneity is not truly unspecific; it is, on the contrary, the result of fine-tuning, optimization, or specialization.
        Without knowing the details of these processes, the resulting diversity appears random or unspecific.
        Moreover, the distinction between type-specific and unspecific forms of heterogeneity relies on the assumption that different neuron types are distinguishable \citep{Battaglia13_13}.
        Without knowing the characteristics separating two neuronal phenotypes, these cell classes are treated as one type, and the observed diversity in neuron and synapse parameters appears unspecific.

        To some extent, synaptic heterogeneity may also result from variations in experimental protocols and in unobserved variables affecting the synapse characteristics.
        Synaptic weights, for example, are often assessed in voltage-clamp experiments as the amplitudes of somatic postsynaptic currents evoked by presynaptic action potentials.
        The resulting synaptic weights are then determined not just by the properties of the pre- and postsynaptic cells or by the synapse type and position but also by the holding potential or the electrical characteristics of the electrode-cell contact.
        Even in the absence of variations in the experimental protocol, the amplitude of the postsynaptic response is affected by fluctuations in the postsynaptic membrane potential and by the pre- and postsynaptic spike history.
        Hidden variables such as the spike history or the synapse position are often not monitored in experimental studies.
        From the modeler's perspective, it is therefore not straightforward to decide what forms of reported heterogeneity should be accounted for in a given model and what forms are perhaps already represented indirectly by other model features (for example, the  voltage dependence of synaptic currents, short-term plasticity, or dendritic filtering in multicompartment models).

        In mathematical models, unspecific heterogeneity is typically accounted for in a probabilistic manner.
        Here, the parameters characterizing synaptic connectivity, such as synaptic weights, time constants, delays, in-degrees, etc., are randomly drawn from certain distributions. In particular, in the brain, many properties follow long-tailed distributions, often approximating the lognormal distribution \citep{Buzsaki2014log, Robinson2021relationships, Morales2022ubiquitous}.
        These distributions, or the parameters characterizing them, such as the mean or the standard deviation, are extracted from experimental data.
        The rationale underlying this probabilistic modeling approach is twofold.
        First, it acknowledges that the synaptic parameters are typically not known for every single synapse in a given network.
        The majority of experimental studies provide data for small subsets of synapses, often pooled across different recording sessions or animals.
        Second, the probabilistic approach greatly simplifies the models, as the total number of parameters is substantially reduced.
        In probabilistic modeling approaches, the ``model'' is not defined by a single instantiation of a network and all its parameters but by the ensemble of many independent realizations generated from a given set of parameter distributions.
        Observations or findings obtained from a single network realization are meaningless unless they appear generically, i.e., frequently, for many different model realizations.

        As described above, synaptic heterogeneity is often the result of an adaptation, development, or fine-tuning process.
        As demonstrated in a number of studies
        (e.g., see \citealp{Prinz-2004_1345,Achard06_e94,Bahuguna17_79}), such processes typically lead to dependencies between parameters.
        Certain plasticity processes, for example, lead to a competition (anti-correlation) between synapses, such that the strengthening of one synapse results in the weakening of another synapse \citep{Abbott00_1178,Tetzlaff11_47}.
        In a comprehensive probabilistic model, the set of parameters $\{\xi_1,\xi_2,\ldots,\xi_n\}$ for a specific network realization is generated   according to a joint probability density function (pdf) $p_{1,\ldots,n}(x_1,x_2,\ldots,x_n)$, describing the probability of observing $\xi_1=x_1$, $\xi_2=x_2$, \ldots, and $\xi_n=x_n$.
        This joint pdf captures all parameter dependencies.
        Many modeling studies neglect parameter dependencies and assume that the joint pdf $p_{1,\ldots,n}(x_1,x_2,\ldots,x_n)=p_1(x_1)p_2(x_2)\ldots{}p_n(x_n)$ factorizes.
        In these studies, each parameter $\xi_i$ is drawn from its respective marginal distribution $p_i(\cdot)$, independently of all other parameters.
        As before, this simplifying assumption typically reflects a lack of knowledge, as the available experimental data generally do not capture parameter dependencies.
        Theoretical studies show that this choice can have detrimental consequences for the dynamical and functional properties of the resulting system.
        \citet{Bahuguna17_79}, for example, demonstrate that when the parameter dependencies are unknown, replacing all parameters by their respective mean (and thereby ignoring diversity) can be a better choice than drawing them from their marginal distributions.

        A more direct approach toward modeling synaptic heterogeneity is the explicit account of known plasticity, learning, or developmental processes that dynamically lead to the observed diversity in synaptic parameters, including the dependencies described above \citep{morrison2007spike,Tetzlaff11_47}.
        Similarly, multivariate parameter distributions may arise from optimization procedures or supervised learning methods fitting the model to some desired target dynamics or behavior \citep{Eliasmith02,Bahuguna17_79,Bellec20_3625}.
        While these top-down approaches are promising and commonly used in state-of-the-art computational neuroscience, they bear the risk that the underlying data or targets do not sufficiently constrain the model of the actual biological system and hence lead to a multitude of solutions that may not be realized in nature.
        A combination of bottom-up and top-down constraints appears to be the most reliable method to reduce this form of uncertainty.

    \subsection{From mathematical models to simulation} \label{sec:heterogeneity_from_models_to_simulation}
        Investigating the role of heterogeneity in synaptic connectivity by means of analytical mathematical methods is challenging \citep{Brunel99,Roxin11_16217}.
        Therefore, theoretical studies often neglect heterogeneity to simplify the mathematical treatment and provide intuitive insight.
        A common strategy underlying many mathematical approaches is to reduce the dimensionality of the neuronal network dynamics by assuming that the network can be decomposed into homogeneous subpopulations, each of which comprises neurons with identical neuronal and synaptic parameters.
        While this approach can account for the specific heterogeneity described above to some extent, it can hardly describe the effects of unspecific heterogeneity.
        Simulation enables us to test whether the insights obtained under these homogeneity assumptions remain valid if heterogeneity in synaptic parameters is considered.

        Even in simulation studies, however, accounting for synaptic heterogeneity is challenging.
        Acknowledging that every synapse is unique requires representing each synapse with the full individual set of parameters.
        In a homogeneous network where all synapses have identical properties, the connectivity is fully described by the adjacency matrix (which neuron is connected to which, and how often) and a small set of parameters describing the synapse characteristics, such as the synaptic weight $w$, the delay $d$, or the synaptic time constant $\tau$.
        In a heterogeneous network where each synapse $\{j\to{}i\}$ is unique, in contrast, the individual weights $w_{ij}$, delays $d_{ij}$, and time constants $\tau_{ij}$ need to be stored for each connection.
        Therefore, representing the heterogeneous connectivity in simulations of neuronal networks at natural density imposes high memory demands for the underlying computing architecture (see \cref{sec:note_limited-weight}).

        In models of neuronal networks with heterogeneous synaptic connectivity, the heterogeneity is either implemented by drawing synapse parameters from predefined distributions or by a self-organization process driven by some plasticity or learning dynamics (see \cref{sec:heterogeneity_from_experiments_to_models}).
        Simulations based on the first, the probabilistic approach, require efficient methods of drawing random numbers from specified distributions during the network generation phase.
        The NEST simulator, for example, permits the high-level specification of probabilistic connection rules by the user (see \cref{sec:connectivity}), including distributions of synaptic weights, synaptic delays, or plasticity parameters.
        The task of generating a specific connectivity realization by drawing random numbers from these distributions is then delegated to fast low-level (C++) routines.
        The second approach relies on simulating plastic networks or on numerical optimization methods.
        Strategies for simulating different forms of synaptic plasticity are described in Sections \ref{sec:structural-plasticity} and \ref{sec:functional-plasticity}.
        Simulating plastic networks with natural connection density is still a major challenge in computational neuroscience.
        Slow biological processes such as learning and development on timescales of hours, days, and years are presently inaccessible to simulation (or restricted to small and highly simplified models) because of the required wall-clock time.
        In this respect, dedicated neuromorphic computing architectures are particularly interesting as simulation platforms for neuroscience, as they offer the potential for faster-than-real-time simulations and hence, for an understanding of plasticity mechanisms on long timescales \citep{Furber16_051001,Wunderlich19_260}.

\section{Notes} \label{sec:notes}

    In this section, we highlight some of the challenges and pitfalls that may be encountered during modeling and simulation.

    \subsection{Keeping model refinements biologically plausible} \label{sec:note_model-refinements}
        Developing a mathematical model that exhibits a dynamic behavior close to empirical biological data involves iterative optimization procedures, e.g., fitting the experimental data or calibrating the model parameters.
        While this optimization improves the model in some aspects, it may, at the same time, alter it in ways no longer motivated by biological observations \citep{Achard06_e94}.
        For example, fitting is inherently biased in that, by definition, it improves the validation of certain model features at the cost of those not included in the fitting procedure. Thus, each optimization step should be conducted cautiously and checked for biological plausibility.

    \subsection{Reproducible simulations} \label{sec:note_reproducibility}
        For small-scale simulations, sometimes custom simulation kernels are written. However, besides possibly duplicating published and established routines,
        these self-made frameworks are likely to contain bugs and lack documentation, for instance, on edge-case behavior. Therefore even for small networks, it helps to use standardized simulators. In particular, these simulators offer the benefit of being well characterized under different operating conditions using a diverse array of automated tests, being updated on a regular release cycle, and benefiting from the open-source model of iterative refinement \citep{zaytsev2013increasing}. All these factors increase the likelihood of long-term reproducible results.

        An individual simulator should exhibit \emph{replicable} behavior: repeated simulations of the same model should yield bitwise identical results, regardless of the number of threads or processing nodes used, due to the use of deterministic pseudo-random number generators. However, simulating the same model on a different platform or using a different numerical solver or time step size for ordinary differential equations (ODEs) may alter the results, especially in network models exhibiting chaotic and unstable dynamics.  Nonetheless, results and conclusions should be \emph{reproducible}, obtaining the same overall quantitative and qualitative conclusions \cite[for a commentary on this terminology, see][]{plesser2018reproducibility}. Reproducibility of results requires the original software to be available (including libraries and other dependencies) and, where applicable, the original (raw or pre-processed) dataset(s) and relevant metadata.

        Similar to the model descriptions, it increases the reproducibility of methods and results \citep{Goodman2016} to use and contribute to existing simulation frameworks by reporting bugs, improving implemented methods, and developing and publishing custom modules of the respective framework, e.g., in NEST, in the form of \emph{extension modules} \refnest{link:ext-module}.

    \subsection{Distribution of compute workload} \label{sec:note_distribution}
        It is beneficial to distribute the workload evenly across compute nodes, even for networks with complex connectivity and heterogeneous population properties. One way to achieve this is a  \emph{round-robin} distribution of neurons across compute nodes, i.e., in the case of $M$ compute nodes, assigning neuron $n$ to node $(n \bmod M)$.

        In general, the computing system's size affects the workload distribution. On small machines, the number of synapses per neuron is larger than the number of compute nodes the simulation runs on. Hence, each neuron typically has many targets on every compute node. However, with growing network size and the emergence of new supercomputer architectures over the last decades, the ratio between the numbers of synapses per neuron and compute nodes is in some cases reversed. On these new-generation supercomputers, the distribution of neurons over many nodes decreases the chance that a neuron shares a node with a connected partner, especially considering the sparsity of biological neuronal networks. Mitigating this issue, even more modern compute nodes follow the opposite trend: they possess more memory and cores per processor and thus more processing power, which reduces the number of nodes required for the simulation and brings the neurons and their targets closer together.

        One of the main computational challenges remains the connectivity of neuronal networks. For example, representing each of the estimated $10^{14}$ synaptic connections in the brain individually by two double-precision numbers requires about 1.6 PB of main memory. Furthermore, neurons form connections with nerve cells not only in their vicinity but also in various remote areas. This feature distinguishes neuronal simulations from simulations of classical physical systems, that use, for example, finite-element methods exploiting the locality of physical interactions. However, memory and communication bandwidth, as well as cache efficiency, are more critical than floating-point performance of spike communication to local and distant targets \citep{vanalbada2014integrating}.

        Future work should address how computer network connectivity and the simulation distribution across compute nodes can follow (simulated) biological network organization, such as a modular organization on both small and large scales in the brain.

    \subsection{Scalability in theory and practice} \label{sec:note_scalability}
        Recording the simulation time under varying network or computing system sizes characterizes an implementation's \emph{scalability}, which is essential to judging its efficiency \citep{Jordan18_2}.
        The scenario of increasing the computing resources while keeping the network size fixed is called \emph{strong scaling}, and the scenario of increasing the network size and computing resources proportionally is called \emph{weak scaling}.

        Ideally, when in strong scaling the hardware becomes twice as powerful, the simulation time is divided by two. However, in practice, the gains are usually less due to communication overhead and other bottlenecks in the system. Sometimes, a parallelization can even exhibit a super-linear speed-up \citep{Kurth2022}, but this behavior might only become apparent at very large computing system scales.

        Scaling up standard network models to test weak scaling can induce unrealistic activity patterns, for example, regarding the regularity and synchrony of spiking. Since synapses tend to vastly outnumber neurons, the number of synapses is an important determinant of the necessary computing resources. Thus, a reasonable approach is to keep the in-degrees constant when increasing the model size and thereby reduce the overall connection probability  \cite[see, e.g.,][]{Jordan18_2}. This method tends to lessen activity correlations between neurons and hence diminish synchrony. In the case of ideal weak scaling, a network of twice the size should run for twice the time, but in practice, the performance is worse due to the same reasons as for strong scaling. Moreover, there is a complex dependence of the scaling behavior on network properties, such as the connectivity's modularity \citep{vanalbada2014integrating}.

    \subsection{Precise spike times in discrete-time simulation} \label{sec:note_precise-spikes}
        A typical simulation of a continuous-time dynamical process runs in discrete steps of time $\Delta t$. However, exchanging events (spikes) on a grid can cause synchronization in the network as a pure simulation artifact. This effect disappears in the limit of $\Delta t \rightarrow 0$, but decreasing the time step increases the time necessary for the simulation to complete, so a tradeoff has to be made. A more computationally efficient solution is to store an extra offset value in spike events, which, in combination with a minimum synaptic delay and an algorithm that finds the precise time of spiking, decouples the simulation time step from the temporal precision with which spikes are exchanged \citep{hanuschkin2010general}.

    \subsection{Simulating until convergence} \label{sec:note_convergence}
        Simulating neuroplasticity for too short of a period, especially longer-timescale processes like normalization and neuromodulation, is a common pitfall.
        Several of these dynamic processes have the potential to cause an abrupt bifurcation in the system late in the simulation.
        Additionally, some plasticity rules produce a long-tailed distribution of synaptic strengths  \citep{FukaiLongTail}, whereby the distribution reaches equilibrium again only after an extended simulation period.
        A limited measurement duration can also be a problem in empirical neuroscience, but simulations can, in principle, run as long as desired.
        The only limitations are the computing resources available (as these need to be shared with other users on high-performance computing systems) and the amount of time the simulation takes to complete, which depends on the simulator's efficiency.
        Simulating until the measure of learning performance has adequately converged can circumvent this pitfall to some degree.

    \subsection{Limited synaptic weight resolution} \label{sec:note_limited-weight}
        Limiting the numerical resolution of synaptic parameters, such as the synaptic weight, appears to be an obvious strategy to reduce memory and compute load.
        To some extent, nature itself copes well with quantized synaptic weights:
        Transmission in chemical synapses is quantized due to the release of neurotransmitters in discrete packages from vesicles in the presynaptic axon terminals.
        The analysis of spontaneous (miniature) postsynaptic currents, i.e., postsynaptic responses to the release of neurotransmitters from single presynaptic vesicles, reveals that the resolution of synaptic weights is indeed finite for chemical synapses.
        As shown by \citet{Malkin14_506}, the amplitudes of spontaneous excitatory postsynaptic currents recorded from different types of excitatory and inhibitory cortical neurons follow a unimodal distribution with a peak at about $20\,\text{pA}$ and a lower bound at about $10\,\text{pA}$.
        Such a cut-off is present despite several factors that may wash out the discreteness of the synaptic transmission, such as variability in vesicle sizes, variability in the position of vesicle fusion zones, quasi-randomness in neurotransmitter diffusion across the synaptic cleft, and variability in postsynaptic receptor densities.
        However, the discreteness of synaptic strengths is obscured for evoked synaptic responses involving neurotransmitter release from many presynaptic vesicles and for superpositions of inputs from many synapses, and thus unlikely to play a particular role in the dynamics of the neuronal network as a whole.

        Inspired by these observations, \citet{Dasbach21_90} systematically investigated the effects of a limited synaptic weight resolution on the dynamics of recurrent spiking neuronal networks resembling local cortical circuits.
        They show that a naive quantization of synaptic weights generally leads to a distortion of the firing statistics.
        However, in the example of one network type, they could demonstrate that the firing statistics remain unaffected under a weight discretization that preserves the mean and variance of the total synaptic input currents.
        In networks with sufficiently heterogeneous in-degrees, the firing statistics stay constant, even when replacing all synaptic weights with the mean of the weight distribution, i.e., entirely neglecting the unspecific form of heterogeneity in synaptic weights.
        Applying this finding in simulations reduces the memory demands substantially.
        The effect of discretized synaptic weights in networks undergoing different forms of synaptic plasticity has rarely been investigated \citep{pfeil20124} and remains a subject for future study.

    \subsection{Precise specification of the model} \label{sec:note_model_specifiction}
        An incomplete and ambiguous description of a model without following conventions can impede understanding by the reader and thus the reproducibility of the respective study.
        The first step to avoiding this pitfall is to gain a clear picture of the requirements a model description must fulfill. For this, \citet{nordlie2009towards} propose that a neuronal network model description must contain a complete and detailed account of its architecture and the dynamics of its parts.
        In the formalization of model descriptions, computational neuroscience is less mature than other fields of science. Nonetheless, best practice guidelines are emerging that suggest the use of standardized tables of the model characteristics that cover the network architecture and connectivity, all neuron and synapse models used, the applied input stimuli, and the recorded data, described by a combination of text, equations, figures, subtables, and pseudocode \citep{nordlie2009towards}. Formalized sketches of the network and unified connectivity concepts \citep{Senk2022} help computational neuroscientists to unequivocally convey their models and, consequently, readers to understand them.

        It is advisable to use one of the numerous formal languages that facilitate such specifications and make them consistent, e.g., NeuroML \citep{neuroml}, NESTML \citep{plotnikov2016nestml}, or PyNN \citep{davison2009pynn}. Most of such languages adhere to the class of either declarative or procedural languages. While a declarative language specifies the model's features, a procedural language specifies the series of commands or instructions needed for constructing the model. Best practices for either of these approaches include formally defined syntax and semantics or an API specification, both uniquely identified by version numbers.

        It is good practice to keep the specification and implementation of a model separate. For example, implementation details such as the time resolution and the spike threshold detection method are essential for the reproducibility of the results but are not part of the model itself. Like the model description, the implementation specification should be complete and sufficient, as it can be challenging to reverse-engineer implementations and test the robustness of the results to implementation alterations \citep{nordlie2009towards}. However, the model should be robust to different choices in such implementation details.

        Generating an executable representation of the model can be fully automated. For example, tools like NESTML specialize in processing the model descriptions of different complexity levels and in verbose formats like XML, creating executable implementations and visualizations, and facilitating debugging \citep{blundell2018code}.

        To summarize, it is advisable to follow standardized model descriptions, implementations, and generation procedures wherever possible for a project and ideally share the models with the community in dedicated databases for computational neuroscience models like ModelDB \citep{McDougal2016} or Open Source Brain \citep{Gleeson2019open}.

\section{Conclusions} \label{sec:conclusions}

    Large-scale neuronal network simulations are a key tool for understanding brain processes. They make the complex nonlinear dynamics of neuronal activity, which are out of reach with analytic methods, accessible to inquiry. Moreover, continuous interaction between computational modeling and advances in empirical understanding have iteratively refined simulation approaches throughout the history of computational neuroscience. We hence outline a few promising directions for empirical and modeling approaches to work together.

    New empirical data can bolster simulation studies in many ways. For example, long-term tracking of synapses \emph{in vivo} will elucidate the relationship between synaptic plasticity and function \citep{holtmaat2009long}. Besides, recordings of complete connectomes at single-neuron resolution are becoming feasible for ever-larger brains and may soon be available on the scale of a mouse brain \citep{foxley2021multi}.
    Data-driven models based on such detailed and specific connectivity are complemented by models whose connectivity is generated with statistical approaches. The latter models also profit from more available data constraining their parameters; higher-level network organization can, for instance, be informed by data on hierarchical modularity and small-world properties.
    Sometimes, the conditions of data retrieval are inconsistent between experiments, or experiments only cover a small subset of the model system, and modeling could benefit from additional studies to fill the gaps.

    How the available data are integrated into models depends on the particular research question and the level of abstraction appropriate for it.
    The modeler needs to decide (or find out) which features and phenomena of the natural system need to be represented in detail and which ones can be approximated.
    For instance, a biophysically detailed model with discrete vesicle release dynamics would be suitable when investigating how a compound influencing vesicle fusion to the membrane affects synaptic transmission. In contrast, investigating the compound's effects on large-scale network dynamics could necessitate approximation of the vesicle release by a simplified set of continuous quantities and differential equations.
    In essence, a good computational model should represent the relevant attributes of the studied biological structure, have explanatory power and simultaneously not necessitate extensive simulation time.

    There are several ways in which the infrastructure of computer simulations itself can innovate, indicated by current trends and feature requests from the community. Simulation efficiency is still a bottleneck, especially in simulations involving plasticity, as they need to run for a comparatively long time. Furthermore, large-scale networks require powerful, high-performance compute clusters, which provide large amounts of RAM, and enable running the simulation in parallel and distributed across many CPUs or GPUs which are interconnected through a low-latency network. More advanced synaptic plasticity rules, for example involving tripartite synapses influenced by astrocytes or neuromodulators like dopamine, still lack software support for efficient simulations on a larger scale. Finally, simulations at cellular resolution could be extended towards multiphysics modeling by incorporating other physical phenomena. Such models could account for the volume diffusion of neuromodulators or the electric field in the neuropil to simulate ephaptic coupling for instance.

    In general, what should one strive for in a model? The statistician George Box famously said, ``All models are wrong, but some are useful.'' Although it helps to remind ourselves of the difficulties of modeling, to say that all models are wrong is not doing them justice, as put by Sir David R.~Cox in a comment on \citet{Chatfield1995}: ``The very word \emph{model} implies simplification and idealization. The idea that complex physical, biological or sociological systems can be exactly described by a few formulae is patently absurd. The construction of idealized representations that capture important stable aspects of such systems is, however, a vital part of general scientific analysis and statistical models \textelp{}''. A model with unlimited parameters can fit the data perfectly but at the cost of generalizability and explanatory power. Ockham's razor, or the law of parsimony, provides a helpful heuristic in this context, guiding us to prefer the theory with fewer parameters between two competing theories. This objective is also formalized in Bayesian information criteria, which penalize models with larger numbers of parameters. In other words, the aim to reduce complexity should guide modeling choices to address a given research question.

    However, condensing the biophysical details and terminology to arrive at a parsimonious, phenomenological formulation could impede testing such a minimal model experimentally.
    Moreover, the rigorous application of Ockham's razor leads to models optimized for single phenomena (e.g., connectivity, synaptic transmission, structural or functional plasticity) that are thus hard to combine. To solve this problem, after reducing the phenomena to their essential variables, we should express each model in a way that allows for their combination.
    Biophysical details could provide contact points between model concepts from different subdomains. Progress in integrating heterogeneous phenomena in large-scale models requires that models act as platforms that can be modified and extended over time. For example, after a minimal model has achieved a satisfactory performance, modelers could test whether the same results still hold for models of greater biophysical detail.

    As a foreseeable trend, more and more computational neuroscientists will adopt procedures ensuring the reproducibility of their methods and results over the following years.
    This approach includes, where possible, data, model, and code sharing in open-access online repositories, adherence to open standards for model formats and software tools, active management of metadata, containerized distribution of dependencies, unit testing, and continuous integration instead of creating new in-house toolchains from the ground up. Last but not least, active contribution to an existing model database or open-source software, be it as small as a feature request, is a low-threshold action conducive to reproducible research everyone can take. Ultimately, the community's research interests in the form of these requests and contributions shape the landscape of available models in an open-access simulator.

\section*{Acknowledgments}
    The authors thank Dennis Terhorst and Jessica Mitchell for technical help with the NEST documentation links, Angela Fischer for advice on the layout of the figures, and Rainer Waser for helpful feedback on the manuscript.
    Furthermore, they gratefully acknowledge funding from the European Union's Horizon 2020 Framework Programme for Research and Innovation under Specific Grant Agreement No. 945539 (Human Brain Project SGA3), from the Joint Lab ``Supercomputing and Modeling for the Human Brain'', the Helmholtz Association Initiative and Networking Fund under project number SO-092 (Advanced Computing Architectures),
    the Federal Ministry of Education and Research BMBF under grant ID 03ZU1106CB within NeuroSys as part of the initiative ``Clusters4Future'',
    the Helmholtz Metadata Collaboration (HMC) under funding ID ZT-I-PF-3-026,
    the Deutsche Forschungsgemeinschaft (DFG, German Research Foundation) under grant ID 368482240/GRK2416,
    HiRSE\_PS, the Helmholtz Platform for Research Software Engineering - Preparatory Study, an innovation pool project of the Helmholtz Association,
    and the Jülich-Aachen Research Alliance Center for Simulation and Data Science (JARA-CSD) School for Simulation and Data Science (SSD),
    as well as computing time granted by the JARA Vergabegremium and provided on the JARA Partition part of the supercomputer JURECA DC at Forschungszentrum Jülich (computation grant JINB33).

\newpage
\section*{References}
\bibliographystyle{abbrvnat}

\begingroup
\renewcommand{\section}[2]{}%

\endgroup

\newpage

\section{Links to the Programs} \label{sec:program-links}
\small
\begin{enumerate}
    \item \url{https://nest-simulator.readthedocs.io/en/latest/tutorials/pynest_tutorial/part_3_connecting_networks_with_synapses.html} \label{link:conn}
    \item \url{https://github.com/nest/nestml/blob/master/models/neurons/wb_cond_multisyn.nestml} \label{link:multisyn-nestml-model}
    \item \url{https://www.nest-simulator.org/py_sample/structural_plasticity/} \label{link:struc-plast}
    \item \url{https://nest-simulator.readthedocs.io/en/latest/auto_examples/tsodyks_depressing.html} \label{link:tsodyks-depress}
    \item \url{https://nest-simulator.readthedocs.io/en/latest/auto_examples/tsodyks_facilitating.html} \label{link:tsodyks-fac}
    \item \url{https://nestml.readthedocs.io/en/latest/models_library/stdp.html} \label{link:stdp}
    \item \url{https://nestml.readthedocs.io/en/latest/nestml_language/synapses_in_nestml.html} \label{link:weight-norm}
    \item \url{https://nest-simulator.readthedocs.io/en/latest/guides/weight_normalization.html} \label{link:weight-normalization}
    \item \url{https://nest-extension-module.readthedocs.io/en/latest/extension_modules.html} \label{link:ext-module}
\end{enumerate} \end{document}